# A Stable Nuclear Future? The Impact of Autonomous Systems and Artificial Intelligence

December 2019

Michael C. Horowitz, Paul Scharre, and Alexander Velez-Green[1]

**Abstract:** The potential for advances in information-age technologies to undermine nuclear deterrence and influence the potential for nuclear escalation represents a critical question for international politics. One challenge is that uncertainty about the trajectory of technologies such as autonomous systems and artificial intelligence (AI) makes assessments difficult. This paper evaluates the relative impact of autonomous systems and artificial intelligence in three areas: nuclear command and control, nuclear delivery platforms and vehicles, and conventional applications of autonomous systems with consequences for nuclear stability. We argue that countries may be more likely to use risky forms of autonomy when they fear that their second-strike capabilities will be undermined. Additionally, the potential deployment of uninhabited, autonomous nuclear delivery platforms and vehicles could raise the prospect for accidents and miscalculation. Conventional military applications of autonomous systems could simultaneously influence nuclear force postures and first-strike stability in previously unanticipated ways. In particular, the need to fight at machine speed and the cognitive risk introduced by automation bias could increase the risk of unintended escalation. Finally, used properly, there should be many applications of more autonomous systems in nuclear operations that can increase reliability, reduce the risk of accidents, and buy more time for decision-makers in a crisis.

[1] Author order is alphabetical. Michael C. Horowitz is Professor of Political Science and Associate Director of Perry World House at the University of Pennsylvania. Paul Scharre is Senior Fellow and Director, Technology and National Security Program at the Center for a New American Security. Alexander Velez-Green is a defense analyst based in Washington, DC. The authors would like to thank the Stanley Foundation for convening a workshop in January 2018 to discuss an early draft of the paper. This article was also made possible, in part, by a grant from Carnegie Corporation of New York, as well as the Air Force Office of Scientific Research and the Minerva Research Initiative under grant #FA9550-18-1-0194. The views expressed here are solely those of the authors. They do reflect and should not be attributed to the policies or positions of the U.S. government. All errors are the responsibilities of the authors.

# Introduction

Nuclear weapons are arguably the single most significant weapon system invented in modern history, meaning uncertainty about the viability of nuclear deterrence in the 21$^{st}$ century constitutes one of the most important security risks facing the world.[2] This uncertainty is both a product and source of increased tensions in nuclear dyads worldwide.

The proliferation of conventional military technologies, such as hypersonic weapons, could further undermine deterrence by potentially undermining traditional modes of escalation management, and as a consequence, nuclear stability.[3] The impact of autonomous systems and artificial intelligence (AI) for nuclear stability remains understudied, however.[4]

In early 2017, Klaus Schwab of the World Economic Forum argued that the world is on the cusp of a Fourth Industrial Revolution, wherein several technologies – but most prominently AI – could reshape global affairs.[5] Many defense experts around the world share Schwab's recognition of the potentially transformative effects of AI.[6] The most prominent statements about the impact of AI on warfare, however, tend to be extreme. Elon Musk, for instance, has vocally contended that AI run amok could risk World War III.[7] This overheated rhetoric masks the way that advances in automation, autonomous systems, and AI may actually influence warfare, especially in the vital areas of nuclear deterrence and warfighting. The intersection of nuclear stability and artificial intelligence thus raises critical issues for the study of international politics.

---

[2] Robert Jervis, *The Meaning of the Nuclear Revolution : Statecraft and the Prospect of Armageddon* (Ithaca: Cornell University Press, 1989).

[3] Keir A Lieber and Daryl G Press, "The New Era of Counterforce: Technological Change and the Future of Nuclear Deterrence," International Security 41, no. 4 (2017): 9-49; Charles L Glaser and Steve Fetter, "Should the United States Reject Mad? Damage Limitation and Us Nuclear Strategy toward China," International Security 41, no. 1 (2016): 49-98; Austin Long and Brendan Rittenhouse Green, "Stalking the Secure Second Strike: Intelligence, Counterforce, and Nuclear Strategy," Journal of Strategic Studies 38, no. 1-2 (2015): 38-73. Also see Brendan Rittenhouse Green et al, "The Limits of Damage Limitation," International Security 42, no. 1 (2017): 193-207.

[4] For exceptions, see Edward Geist and Andrew J. Lohn, "How Might Artificial Intelligence Affect the Risk of Nuclear War," *RAND Corporation*, 2018, https://www.rand.org/pubs/perspectives/PE296.html; Vincent Boulanin and Maaike Verbruggen, "Mapping the Development of Autonomy in Weapon Systems," *Stockholm International Peace Research Institute*, November 2017, https://www.sipri.org/publications/2017/other-publications/mapping-development-autonomy-weapon-systems; Technology for Global Security, "Artificial Intelligence and the Military: Forever Altering Strategic Stability," *T4GS Report*, February 13, 2019, https://www.tech4gs.org/uploads/1/1/1/5/111521085/ai_and_the_military_forever_altering_strategic_stability__t4gs_research_paper.pdf.

[5] Klaus Schwab, The Fourth Industrial Revolution (New York: Crown Business, 2017).

[6] For examples, see Paul Scharre, *Army of None* (New York: W.W. Norton & Company, 2018); Michael C. Horowitz, "Artificial Intelligence, International Competition, and the Balance of Power," *Texas National Security Review*, vol. 3, no 1. (2018), pp. 38-57; Peter Asaro, "Why the World Needs to Regulate Autonomous Weapons, and Soon," *Bulletin of the Atomic Scientists*, April 27, 2018, https://thebulletin.org/2018/04/why-the-world-needs-to-regulate-autonomous-weapons-and-soon/; Heather M. Roff and Richard Moyes. (2016). "Meaningful human control, artificial intelligence and autonomous weapons," in Briefing Paper Prepared for the Informal Meeting of Experts on Lethal Autonomous Weapons Systems, UN Convention on Certain Conventional Weapons, http://www.article36.org/wp-content/uploads/2016/04/MHC-AI-and-AWS-FINAL.pdf.

[7] Seth Fiegerman, "Elon Musk Predicts World War III," Cnn.com, September 4 2017, http://money.cnn.com/2017/2009/2004/technology/culture/elon-musk-ai-world-war/index.html.

Relative peace between nuclear-armed states in the 20th century arguably relied in part on mutually assured destruction (MAD).[8] MAD prevails when each side recognizes that both it and its opponent have an assured nuclear second-strike capability, or that either side can impose unacceptable damage on the other in retaliation against a nuclear attack.[9] Threat of mutual destruction ultimately led both the United States and the Soviet Union to deprioritize the role of preemption in their nuclear war plans.[10]

Furthermore, as Albert Wohlstetter found, the threat of mutual destruction "offer[ed] every inducement to both powers to reduce the chance of accidental war."[11] While there are no known instances of accidental war, there are historical examples of unintended escalation, either in pre-conflict crises or once a conflict is underway.[12] Accidental escalation is when a state *unintentionally* commits an escalatory act (i.e. due to technical malfunction, human error, or incomplete control over military forces).[13] Inadvertent escalation can also occur, whereby a state *unknowingly* commits an escalatory act (i.e., an intentional act that unknowingly crossing an adversary's red line).[14] Accidents have increased tensions between countries on numerous occasions, but have not led to escalation.[15] Nuclear-armed states have expended vast resources to minimize the risk of unintentional escalation, knowing that it could lead to catastrophe should it occur.

Automation may complicate the risks of escalation, deliberate or unintended, in a number of ways. Automation has improved safety and reliability in other settings, from nuclear power plants to commercial airliners. Used properly, many applications of automation in nuclear operations could increase reliability, reduce the risk of accidents, and buy more time for decision-makers in a crisis. Automation can help ensure that information is quickly processed, national leaders' desires are swiftly and efficiently conveyed, and launch orders are faithfully executed.

---

[8] "Nuclear stability" is a subcomponent of the broader concept of "strategic stability." For a history of both, see Elbridge A. Colby and Michael S. Gerson, eds, S*trategic Stability: Contending Interpretations* (Carlisle Barracks, P.A.: Strategic Studies Institute and the U.S. Army War College, February 2013).
[9] As Thomas Schelling wrote, "The balance is stable only when neither, in striking first, can destroy the other's ability to strike back." Thomas C. Schelling, *The Strategy of Conflict* (Cambridge, M.A.: Harvard University Press, 1960), 232. See also Thomas C. Schelling, *Arms and Influence* (New Haven, C.T.: Yale University Press, 1966), 228-229.
[10] Alexander Velez-Green, "The Unsettling View from Moscow: Russia's Strategic Debate on a Doctrine of Preemption" (Center for a New American Security, April 2017). Technology for Global Stability, "Artificial Intelligence and the Military: Forever Altering Strategic Stability," https://www.tech4gs.org/uploads/1/1/1/5/111521085/ai_and_the_military_forever_altering_strategic_stability__t4gs_research_paper.pdf..
[11] Albert Wohlstetter, "The Delicate Balance of Terror," *Foreign Affairs* (January 1959); and Albert Wohlstetter, *The Delicate Balance of Terror*, P-1472 (RAND Corporation, December 1958).
[12] Forrest E. Morgan et al., *Dangerous Thresholds: Managing Escalation in the 21st Century* (RAND Corporation, 2008) 23-28.
[13] Morgan et al., *Dangerous Thresholds,* 25-28; and Scott D. Sagan, *The Limits of Safety : Organizations, Accidents, and Nuclear Weapons* (Princeton, N.J.: Princeton University Press, 1993). On the stability of crises to unintended escalation, see Dan Reiter, "Exploding the Powder Keg Myth: Preemptive Wars Almost Never Happen," *International Security*, Vol. 20, No. 2 (1995), pp. 5-34.
[14] Kerry M. Kartchner and Michael S. Gerson, "Escalation to Limited Nuclear War in the 21st Century," in *On Limited Nuclear War in the 21st Century*, eds. Jeffrey A. Larsen and Kerry M. Kartchner (Stanford, C.A.: Stanford University Press, 2014), 150, 158. See also Bruce G. Blair, *The Logic of Accidental Nuclear War* (Washington, D.C, Brookings Institution Press, 1993); Scott D. Sagan, *The Limits of Safety: Organizations, Accidents, and Nuclear Weapons* (Princeton, N.J.: Princeton University Press, 1993); Barry R. Posen, *Inadvertent Escalation: Conventional War and Nuclear Risks* (Ithaca, N.Y.: Cornell University Press, 1991); Morgan et al., *Dangerous Thresholds,* 23-25.
[15] Patricia Lewis et al., "Too Close for Comfort: Cases of Near Nuclear Use and Options for Policy," Chatham House Report, April 2014.

On the other hand, poor applications of automation could render nuclear early warning or command-and-control (C2) systems more opaque to users, leading to human-machine interaction failures. Human users could fall victim to automation bias, for example, surrendering their judgment to the system in a crisis.

Automation is often brittle and lacks the flexibility humans have to react to events in their broader context. The states most likely to be willing to tolerate these risks for the perceived capability gains would be those that have significant concerns about the viability of their second strike deterrents. Thus, the more a country fears that, in a world without using autonomous systems, its ability to retaliate to a nuclear strike would be at risk, the more attractive autonomous systems may appear.

Uninhabited nuclear delivery platforms could undermine nuclear surety, as they could be hacked or slip out of control, potentially leading to accidental or inadvertent escalation. Automated systems could end up reducing decision-maker flexibility by narrowing options, hampering attempts to manage escalation.

These dynamics suggest that autonomous systems could influence the potential for nuclear escalation in three ways. First, while many aspects of the nuclear enterprise are already automated in many countries, from early warning and command and control to missile targeting, as autonomous systems improve, states may elect to automate new portions of the early warning and C2 processes to improve both performance and security. From a security standpoint, for instance, increased automation in nuclear early warning may allow operators to identify threats more rapidly in a complex environment. Likewise, automation may help to ensure the dissemination of launch orders in a timely manner in a degraded communications environment. States may also automate – or threaten to automate – nuclear launch procedures in the belief that doing so would provide them with a coercive advantage over adversaries.

Second, as military robotics advance, nuclear powers could deploy uninhabited nuclear delivery platforms for a variety of reasons. For instance, a state might deploy nuclear-armed long-endurance uninhabited aerial vehicles (UAVs) in the belief that doing so would provide additional nuclear signaling or strike options. They might also look to uninhabited nuclear delivery platforms to bolster their secure second-strike capabilities. Nuclear delivery vehicles – such as torpedoes – capable of autonomously countering enemy defenses or selecting targets might be seen to do likewise. Alternatively, a government might choose to automate its nuclear forces so that a small number of trusted agents can maintain control. This might could be especially attractive for a nuclear-armed regime that fears a coup or other forms of interference by its nation's military elite.

Third, the increased automation of *conventional* military systems might influence nuclear stability in direct and indirect ways.[16] It may enable – or more likely yet, be seen to enable – improved counterforce operations by technologically-advanced states. The ineffectiveness of counterforce operations – and hence the survivability of second-strike deterrents – presently hinges in large part on the difficulty of finding and tracking adversary nuclear launch platforms (mobile missiles or submarines) long enough for ordnance to be delivered. Machine learning algorithms and other applications of artificial intelligence could, in principle, improve states' abilities to collect and sift through large amounts of data in order to locate and track such targets, though it is important to recognize limitations to any developments given the real-time requirements for a disarming strike. Likewise, military autonomy could enable the deployment of conventional autonomous systems designed to shadow and/or attack nuclear-armed submarines. Furthermore, if automation gives (or is perceived to give) one side in a competitive dyad a significant conventional military advantage, the weaker side may feel compelled to rely more heavily on nuclear weapons for deterrence and warfighting.

These issues surrounding the potential impacts of artificial intelligence are magnified by uncertainty about the trajectory of technological developments. This article first proceeds by clarifying what autonomous systems are and clarifying often-tricky definitional issues surrounding artificial intelligence. It then lays out some key theoretical expectations. Second, the article explores the impact of autonomous systems on early warning and nuclear command and control, as well as intelligence, surveillance, and reconnaissance (ISR) relevant for nuclear systems, in the context of recent research. Third, the article discusses the potential for uninhabited nuclear delivery platforms and vehicles featuring new kinds of automation. Fourth, the article describes the way conventional autonomous systems could both directly and indirectly influence nuclear stability. Finally, the article concludes by assessing the net likely impact of autonomous systems on nuclear stability and describing potential pathways for future research. The analysis argues that the impact of autonomous systems could depend on the specific application – both where automation falls in the nuclear enterprise but also how it is implemented in terms of design, human-machine interfaces, training, and operator culture.

## Autonomous Systems and Artificial Intelligence

The field of artificial intelligence, which dates back to the 1950s, has seen tremendous growth in recent years. Much of these recent gains have come from "deep learning," a machine learning technique that uses deep (multi-layer) neural networks.

Deep learning is relatively new and, while a powerful technique, has certain insecurities. Deep neural networks are vulnerable to a form of spoofing attack that uses adversarial data to fool the network into misidentifying false data with high confidence.[17] This vulnerability is prevalent across neural networks in wide use today. While adversarial training can somewhat mitigate these risks, there is currently no known solution to this vulnerability.[18] Additionally, machine learning is vulnerable to "data poisoning" techniques that manipulate the data used to train a machine learning system, thus causing it to learn the wrong thing. Finally, artificial intelligence systems today, including those that do not use deep learning, have a set of safety challenges broadly referred to as "the control problem."[19] Under certain conditions, for example, artificial intelligence tools can learn in unexpected and counterintuitive ways that may not be consistent with their users or designer's intent.[20]

---

[16] Paul Scharre and Michael C. Horowitz, "An Introduction to Autonomy in Weapon Systems," CNAS Working Paper, February 2015, http://www.cnas.org/sites/default/files/publications-pdf/Ethical%20Autonomy%20Working%20Paper_021015_v021002.pdf.

[17] Ian J Goodfellow, Jonathon Shlens, and Christian Szegedy, "Explaining and Harnessing Adversarial Examples," *arXiv preprint arXiv:1412.6572* (2014); Anh Nguyen, Jason Yosinski, and Jeff Clune, "Deep Neural Networks Are Easily Fooled: High Confidence Predictions for Unrecognizable Images" (paper presented at the Proceedings of the IEEE Conference on Computer Vision and Pattern Recognition, 2015); Christian Szegedy et al, "Intriguing Properties of Neural Networks," *arXiv preprint arXiv:1312.6199* (2013).

These machine learning tools are powerful and are being used in novel experimental applications.[21] But, given these vulnerabilities, they are not sufficiently mature to operate independent of human control for high-risk tasks, such as nuclear operations. These vulnerabilities are fairly well-understood in AI circles and among technical experts within the U.S. defense community.[22] As a result, while AI and machine learning tools are already being incorporated into a variety of commercial applications,[23] it seems unlikely that risk-averse government bureaucracies will be at the forefront of adoption, particularly for high-risk applications such as nuclear operations.

However, older "first wave" AI systems that employ rule-based decision-making logic have been used in automated and autonomous systems for decades, including in nuclear operations.[24] These expert AI systems use handcrafted knowledge from humans to create a structured set of if-then rules to determine the appropriate action in a given setting. Automated systems of this type are widely used, including in high-consequence operations such as commercial airline autopilots and automation in nuclear power plant operations. Rule-based expert AI systems can often improve reliability and performance when used in predictable settings. However, because such systems can only follow the rules they've been given, they often perform poorly in novel situations or unpredictable environments.

In this paper, unless otherwise specified, we generally use the terms automated or autonomous system to refer to "first wave" expert AI systems that perform various tasks on their own, sometimes under human supervision (supervised autonomy) and sometimes absent human supervision for a period of time (full autonomy).

---

[18] Ian Goodfellow, "Deep Learning Adversarial Examples Clarifying Misconceptions," *KDnuggets*, July 2015, http://www.kdnuggets.com/2015/2007/deep-learning-adversarial-examples-misconceptions.html.
[19] Dario Amodei et al, "Concrete Problems in Ai Safety," *arXiv preprint arXiv:1606.06565* (2016).
[20] Amodei et al, "Concrete Problems in AI Safety;" Joel Lehman et al., "The Surprising Creativity of Digital Evolution," August 14, 2018, https://arxiv.org/pdf/1803.03453.pdf.
[21] "AlphaGo," https://deepmind.com/research/alphago/. Also see Mariusz Bojarski et al, "End to End Learning for Self-Driving Cars," *arXiv preprint arXiv:1604.07316* (2016); Volodymyr Mnih et al, "Playing Atari with Deep Reinforcement Learning," *arXiv preprint arXiv:1312.5602* (2013).
[22] JASON Study, "Perspectives on Research in Artificial Intelligence and Artificial General Intelligence Relevant to DoD," *The MITRE Corporation, JSR-16-Task-003*, January 2017, https://fas.org/irp/agency/dod/jason/ai-dod.pdf; U.S. Air Force Office of the Chief Scientist, "Autonomous Horizons: System Autonomy in the Air Force – a Path to the Future, Volume I: Human-Autonomy Teaming," *AF/ST TR 15-01 June*, https://www.hsdl.org/?abstract&did=768107..
[23] "Watson," https://www.ibm.com/watson/. H. James Wilson, Paul Daugherty, and Prashant Shukla, "How One Clothing Company Blends Ai and Human Expertise," *Harvard Business Review*, November 21 2016, https://hbr.org/2016/2011/how-one-clothing-company-blends-ai-and-human-expertis.
[24] John Launchbury, "DARPA Perspective on AI," Defense Advanced Research Projects Agency, https://www.darpa.mil/about-us/darpa-perspective-on-ai.

*To Trust or Not to Trust: Autonomous Systems*

Automation has been used in high-risk applications for decades, in both civilian and military capacities. Nuclear power plants, commercial airlines, and private space ventures, for instance, all use automation to perform complex operations. Automation also serves niche roles in nuclear operations, including in early warning, targeting, launch control, delivery platforms, and delivery vehicles. Each of these applications, however, relies on mature technology and often retains human control over decision-making prior to the launch of the delivery vehicle.

Questions about adopting autonomous systems require potential adopters to grapple with how to balance the risk that humans will not trust machines to operate effectively against the risk that humans will trust machines too much. Trust gaps occur when people do not trust machines to do the work of people, even if the machine outperforms humans in benchmark tasks. This can lead to an unwillingness to deploy or properly use systems. It can also lead to a preference for using human-controlled systems. Dietvorst, Simmons, and Massey show that when humans have to choose between using human forecasters or algorithms to make assessments about the future, they trust humans even when they can see an algorithm outperform humans.[25] Moreover, when algorithms make mistakes, humans are faster to lose confidence in their effectiveness than when they see humans make mistakes.

There is contested evidence that a trust gap exists when it comes to military remotely-piloted aircraft (i.e., drones). Surveying Ground Tactical Air Controllers (GTACs) about their preference for inhabited aircraft versus drones for close air support, Schneider and MacDonald find that GTACs tended to prefer inhabited aircraft. They argue that GTACs believed that pilots in inhabited aircraft had more "skin in the game" and were thus more likely to perform effectively (even though there was no evidence that that was the case).[26] Their results are controversial – other military personnel argue that their trust gap findings do not reflect the reality on the ground, where military personnel are learning to trust remotely-piloted systems [27] Regardless of whether a trust gap exists in the GTAC community, the theoretical point has relevance for thinking about potential end users of autonomous systems.

For applications of artificial intelligence, the alternative to a trust gap is automation bias. While humans are slow to trust information from a "machine," such as data from radar, research demonstrates that once they believe in the effectiveness of the system, they become more willing to surrender judgment, even when there is evidence that the machine may be incorrect in a given situation. For example, in flight simulation experiments, participants given very good, but not perfect, automated flight systems tend to make more mistakes. Participants became more likely to miss problems unless explicitly prompted by the autonomous system; they also tend to trust the autonomous system over their own judgment even when their training suggested the plane might be at risk (e.g. errors of both omission and commission).[28]

---

[25] Berkeley J. Dietvorst, Joseph P. Simmons, and Cade Massey, "Algorithm Aversion: People Erroneously Avoid Algorithms After Seeing Them Err," *Journal of Experimental Psychology: General*, 144, no. 1 (2015): 114-126; Berkeley J. Dietvorst, Joseph P. Simmons, and Cade Massey, "Overcoming Algorithm Aversion: People will Use Imperfect Algorithms if they can (even slightly) modify them," *Management Science*, 64, no. 3: 1155-1170.
[26] Jacquelyn Schneider and Julia MacDonald, "The Role of Risk in Battlefield Force Employment Decisions: An Analysis of Tactical Decisions to Use Unmanned versus Manned Weaponry," *Security Studies*, forthcoming.
[27] Cory T. Anderson et al., "Trust, Troops, and Reapers: Getting 'Drone Research' Right," *War on the Rocks*, April 3, 2018, https://warontherocks.com/2018/04/trust-troops-and-reapers-getting-drone-research-right/

Automation bias, whereby humans effectively surrender judgment to machines, therefore represents one important risk from automation. For example, Army investigators found that automation bias was a factor in the 2003 Patriot fratricides, in which Army Patriot air and missile defense operators shot down two friendly aircraft during the opening stages of the Iraq War. In both instances, humans were "in the loop" and retained final decision authority for firing, but operators nevertheless trusted the (incorrect) signals they were receiving from their automated radar systems.[29] Army investigators found that automation bias was pervasive throughout the Patriot community at the time, which had a culture of "trusting the system without question."[30] According to Army researchers, Patriot operators exhibited an "unwarranted and uncritical trust in automation. In essence, control responsibility [was] ceded to the machine."[31]

In addition to the problems of trust gap and automation bias, human-machine interaction failures can manifest in other ways. The opacity of complex machines can be a hurdle to operators fully understanding them, and this can lead to accidents. As systems increase in complexity, human users may not fully comprehend how automated systems will behave in response to certain inputs coming either from the environment or from human operators themselves.[32] This complexity and breakdown in human-machine integration appears to have been a factor in the 2016 fatal accident involving a Tesla car on autopilot[33] and the 2009 crash of Air France Flight 447, which killed everyone onboard.[34] In both cases, the human users failed to understand how their respective automated systems would respond to their environments, leading them to take actions that, had they known otherwise, they would likely not have taken. A breakdown in human-machine integration can have disastrous consequences even if human users retain manual control over the system, as they did in the case of the Air France 447 crash.

---

[28] Linda J Skitka, Kathleen L Mosier, and Mark Burdick, "Does Automation Bias Decision-Making?," *International Journal of Human-Computer Studies* 51, no. 5 (1999): 991-1006; Mary L Cummings, "Automation Bias in Intelligent Time Critical Decision Support Systems" (paper presented at the AIAA 1st Intelligent Systems Technical Conference, 2004).
[29] John K. Hawley, "Patriot Wars: Automation and the Patriot Air and Missile Defense System," *Center for a A New American Security*, January 25 2017, https://www.cnas.org/publications/reports/patriot-wars; John K. Hawley, "Not by Widgets Alone: The Human Challenge of Technology-Intensive Military Systems," *Armed Forces Journal* 41 (2011): 24-28; John K. Hawley, "Looking Back at 20 Years of Manprint on Patriot: Observations and Lessons," *Army Research Laboratory, ARL-SR-0158*, September 2007, http://www.arl.army.mil/arlreports/2007/ARL-SR-0158.pdf.
[30] Hawley, "Looking Back at 20 Years of Manprint on Patriot."
[31] Hawley, "Not by Widgets Alone." Patriot operators now train on this and other similar scenarios to avoid this problem of unwarranted trust in the automation.
[32] Paul Scharre, "Autonomous Weapons and Operational Risk," *CNAS Working Paper*, February 2016, http://www.cnas.org/sites/default/files/publications-pdf/CNAS_Autonomous-weapons-operational-risk.pdf.
[33] The Tesla Team, "A Tragic Loss," *Tesla.com*, June 30 2016, https://www.tesla.com/blog/tragic-loss; Anjali Singhvi and Karl Russell, "Inside the Self-Driving Tesla Fatal Accident," *The New York Times*, July 12 2016, https://www.nytimes.com/interactive/2016/2007/2001/business/inside-tesla-accident.html.
[34] The full, official accident report by French authorities is, "Final Report: On the accidents of 1st June 2009 to the Airbus A330-203 registered F-GZCP operated by Air France flight 447 Rio de Janeiro – Paris," Bureau d'Enquêtes et d'Analyses pour la sécurité de l'aviation civile, [English translation], 2012, http://www.bea.aero/docspa/2009/f-cp090601.en/pdf/f-cp090601.en.pdf.

Finally, automated systems can pose risks because of their complexity, tight-coupling, and ability to take actions at machine speed. Complex automated systems are generally powered by software, making them potentially vulnerable to bugs and hacking. As one example, a 2007 software malfunction caused eight F-22 fighter jets to lose navigation, fuel subsystems, and some communications when they crossed the international dateline.[35] Rigorous software testing can reduce error rates, but error-free software is not realistic outside of extremely narrow applications.[36] Software vulnerabilities can also leave the door open for hackers. Security researchers have demonstrated the ability to remotely hack automobiles, for example, disabling or taking control of critical driving functions such as steering and brakes.[37] Autonomous systems are also vulnerable to so-called "normal accidents" arising from the interaction of components of complex systems. While these risks are not unique to automation – normal accidents occur in manual systems[38] – automation can increase the "tight coupling" between components. Tight coupling is a condition in which one action in a system directly causes another action with little "slack" in the system – time, visibility, and opportunity for human intervention to manage unanticipated events. Automation can increase the coupling between components and, moreover, accelerate the pace of actions to machine speed. The "brittleness" inherent to emerging forms of automation, along with omnipresent risk of automation bias, human-machine interaction failures, and unanticipated machine behavior, all potentially limit the roles that automation can safely fill.

*Risk, Reliability, and Safety*

There are many models for coping with these risks. One model is to eschew automation entirely, which forgoes its potential benefits. Another model is to retain humans either "in the loop" of automation, requiring positive human input, or "on the loop," overseeing the system in a supervisory capacity. Human-machine teaming is no panacea for these risks. Even if automated systems have little "autonomy" and humans remain in control, the human users could fall victim to automation bias, leading them to cede judgment to the machine. Appropriately determining when to intervene is not sufficient to ensure safe operation. Human users may lack the ability to effectively intervene and take control if their skills have atrophied due to relying on automation or if they do not sufficiently understand the automation.[39] Properly integrating humans and machines into effective and reliable joint cognitive systems that harness the advantages of each requires proper design, testing, training, leadership, and culture, which is not straightforward to achieve.[40]

---

[35] Remarks by Air Force retired Major General Don Sheppard on "This Week at War," *CNN,* February 24, 2007, http://transcripts.cnn.com/TRANSCRIPTS/0702/24/tww.01.html.
[36] Steve McConnell, *Code Complete: A Practical Handbook of Software Construction* (Redmond, WA: Microsoft Press, 1993). The space shuttle is an interesting exception. NASA drove errors to zero through a labor-intensive process employing teams of engineers. However, the space shuttle has only approximately 500,000 lines of code. The F-35 Joint Strike Fighter, in contrast, has over 20 million lines of code. Charles Fishman, "They Write the Right Stuff," FastCompany.com*, December 31, 1996, http://www.fastcompany.com/28121/they-write-right-stuff.
[37] Andy Greenberg, "Hackers Remotely Kill a Jeep on the Highway – With Me in It," *Wired*, July 21, 2015, http://www.wired.com/2015/07/hackers-remotely-kill-jeep-highway/.
[38] Charles Perrow, *Normal Accidents: Living with High Risk Systems* (New York: Basic Books, 1984).. Also see "Eastern L-1011 in Florida: Accident Overview," Federal Aviation Administration, http://lessonslearned.faa.gov/ll_main.cfm?TabID=3&LLID=8&LLTypeID=2#null.
[39] Konnikova, "The Hazards of Going on Autopilot; Stephen M Casner et al, "The Retention of Manual Flying Skills in the Automated Cockpit," *Human factors* 56, no. 8 (2014): 1506-1516.

Reliability and predictability are significant factors in determining whether automation is a net positive or negative in high-risk applications. Constructing reliable joint cognitive systems may be especially difficult for military organizations. Research suggests that organizations' ability to accurately assess risks and deploy reliable systems depends as much, if not more, on bureaucratic and organizational factors as on technical ones.[41] It is not sufficient for safe and reliable operation to be technically possible. Militaries must also be willing to pay the cost – in time, money, and operational effectiveness – of investing in safety. These are challenges for any high-risk applications of automation, but nuclear operations pose certain unique challenges. The potentially cataclysmic consequences of nuclear war makes it difficult to estimate how safe is safe enough.

*Competition and Autonomous Systems*

A challenge to safety in military settings is that operations occur in a competitive environment. Unlike in other areas where safety is paramount, such as airline travel or nuclear power plant operations, in military settings, safety is balanced against operational effectiveness. For nuclear operations, this balancing is captured in the "always-never dilemma."[42] Nuclear organizations are expected to *always* be ready to launch a nuclear weapon at a moment's notice and, at the same time, *never* allow unauthorized or accidental launch. As Scott Sagan points out, meeting both criteria is effectively "impossible."[43]

On one level, the obvious destructive potential of nuclear weapons naturally induces caution in military professionals and policymakers who may be considering whether or not to use automation in nuclear operations. In this sense, a strong organizational bias towards maintaining positive human control over nuclear weapons is likely to mitigate against any risks from adding automation. The track record of safety for nuclear operations might lead one to be less sanguine about the ability of bureaucracies to successfully manage the risks of automation, however.[44]

---

[40] Hawley, "Patriot Wars."
[41] On automation and risks in military decision-making, see ibid; Hawley, "Not by Widgets Alone. On general applications to militaries, see Barry R. Posen, *The Sources of Military Doctrine: France, Britain, and Germany between the World Wars* (Ithaca: Cornell University Press, 1984); Michael C. Horowitz, *The Diffusion of Military Power: Causes and Consequences for International Politics* (Princeton, NJ: Princeton University Press, 2010); Sagan, *The Limits of Safety*; Stephen P. Rosen, *Winning the Next War : Innovation and the Modern Military* (Ithaca: Cornell University Press, 1991); Dima Adamsky, *The Culture of Military Innovation: The Impact of Cultural Factors on the Revolution in Military Affairs in Russia, the Us, and Israel* (Stanford, CA: Stanford University Press, 2010). On organizational reliability and accidents in general, see Perrow, *Normal Accidents: Living with High Risk Systems*; Karl E Weick and Kathleen M Sutcliffe, *Managing the Unexpected: Resilient Performance in an Age of Uncertainty*, vol. 8 (New York: John Wiley & Sons, 2011).
[42] Sagan, *The Limits of Safety*, 278
[43] Ibid.
[44] Sagan, *The Limits of Safety : Organizations, Accidents, and Nuclear Weapons*; Patricia M Lewis et al, "Too Close for Comfort: Cases of near Nuclear Use and Options for Policy," *Chatham House Report*, April 2014, https://www.chathamhouse.org/sites/files/chathamhouse/field/field_document/20140428TooCloseforComfortNuclearUseLewisWilliamsPelopidasAghlani.pdf.

Examples like the Soviet Perimeter system, discussed below, also demonstrate that some nations are likely to view the risks and benefits of automation differently from others.

What might lead to variation in how countries make choices about the relative utility of autonomous systems? The answer could depend on how secure they feel about their non-autonomous nuclear systems. States that feel extremely secure in their second strike capabilities at present may see fewer advantages to automation. In that case, the advantages greater speed and precision but might not appear worth the potential risk of accidents. Instead, states like the United States would likely prefer to use existing non-autonomous systems for nuclear command and control and delivery.

In contrast, countries whose nuclear arsenals are more insecure may be more accepting of risk and may find the perceived advantages of automation more valuable. If a country thinks that its nuclear command and control might be at risk of severe degradation or destruction, it might be more likely to automate early warning to increase its response speed, deploy autonomous nuclear delivery platforms with higher endurance, automate new aspects of target selection for nuclear delivery vehicles, or shift towards more automated nuclear launch postures. All may not happen in unison, of course, but as a general relationship, countries whose arsenals are more insecure may be more willing to take risks to better enhance their arsenal's survivability.

## Autonomous Systems and Nuclear Stability

This section outlines the three areas described in the introduction of the article – nuclear early warning and command and control, nuclear delivery platforms and vehicles, and non-nuclear autonomous systems – and how they could influence nuclear stability, particularly in crisis situations.

### *Automation in Early Warning and Nuclear Command and Control*

There are many places where automation is already used in early warning and nuclear command and control. Early warning systems rely heavily on automation to quickly warn human operators about potential inbound missiles.[45] Command-and-control systems also have a number of automated functions, such as rapid retargeting or communication rockets to beam launch codes down to missile silos.[46] Some forms of automation in early warning and nuclear command and control have been non-controversial. Others have generated significant controversy and have even been involved in near-accidents.

Risk of Catastrophic Failure

---

[45] Geoffrey Forden, Pavel Podvig, and Theodore A Postol, "False Alarm, Nuclear Danger," *IEEE Spectrum* 37, no. 3 (2000): 31-39; Air Force Space Command, "Ballistic Missile Early Warning System," March 2017, http://www.afspc.af.mil/About-Us/Fact-Sheets/Display/Article/1126401/ballistic-missile-early-warning-system; John Pike, "Russia and Early Warning Systems," *GlobalSecurity.org*, November 12 2016, http://www.globalsecurity.org/space/world/russia/warning.htm.
[46] Interview with Gen.-Col. (Ret.) Varfolomei Vladimirovich Korobushin, December 10, 1992, http://nsarchive.gwu.edu/nukevault/ebb285/vol%20II%20Korobushin.PDF; and Interview with Vitalii Leonidovich Kataev, May 1993, http://nsarchive.gwu.edu/nukevault/ebb285/vol%20II%20Kataev.PDF.

The most famous automation-related near-accident is the 1983 Stanislav Petrov incident in which the Soviet Oko satellite-based early warning system registered a false alarm of five U.S. intercontinental ballistic missile (ICBM) launches. Lieutenant Colonel Stanislav Petrov was the watch officer on duty who was responsible for alerting Soviet leadership of a U.S. attack. Petrov has said that the automated alert system reported a missile strike with the "highest" confidence. Automated alerts included an audible siren and a large red backlit screen that flashed "launch" and "missile strike." While these automated alerts serve the purpose of gaining human operators' attention, they can also exacerbate the risk of automation bias. Petrov subsequently reported that he "had a funny feeling in my gut" and estimated the odds of the strike being real as 50/50.[47] Petrov did not fall victim to automation bias and reported a system malfunction to his superiors, rather than reporting a U.S. nuclear strike was underway.

A related highly-controversial application of automation in NC2 is the use of "dead hand" systems that could launch a nuclear counterattack in the event that a nation's leadership was wiped out by a surprise nuclear first strike. The concept of a dead hand "doomsday machine" was a plot point in Stanley Kubrick's 1964 film *Dr. Strangelove*, but there are reports that the Soviet Union may have built a semi-automated dead hand system called "Perimeter." According to primary source interviews with former Soviet officers, the system was intended to be activated in a crisis[48] as a "fail deadly" mechanism to ensure retaliation against the United States in the event of a successful U.S. first strike.

Specific accounts of Perimeter's functionality differ,[49] but the essential concept was a network of light, radiation, seismic, and pressure sensors that could detect any nuclear detonations in Soviet territory.[50] According to accounts of the system, if it was activated and it sensed a nuclear detonation, it would check for communications to the Soviet military General Staff.[51] If there was no order to halt, after a pre-determined period of time ranging from on the order of 15 minutes to an hour, the system would act like a "dead man's switch" and transfer nuclear launch authority directly to human duty officers in an underground bunker. These individuals would then be tasked with launching communication rockets that would fly over Soviet territory, beaming down launch codes to hardened missile silos.[52] A human would remain "in the loop" for the final decision to launch, but Perimeter would bypass normal layers of command and transfer authority directly to the watch officer on duty.[53]

---

[47] Pavel Aksenov, "Stanislav Petrov: The man who may have saved the world," BBC News, September 26, 2013, http://www.bbc.com/news/world-europe-24280831; and David Hoffman, "'I Had A Funny Feeling In My Gut," *The Washington Post*, February 10, 1999, http://www.washingtonpost.com/wp-srv/inatl/longterm/coldwar/shatter021099b.htm.

[48] Interview with Gen.-Col. (Ret.) Varfolomei Vladimirovich Korobushin, 107; Bruce G. Blair, "Russia's Doomsday Machine," *The New York Times*, October 8, 1993, http://www.globalzero.org/files/bb_russias_doomsday_machine_10.08.1998.pdf; and Nicholas Thompson, "Inside the Apocalyptic Soviet Doomsday Machine," *WIRED*, September 1, 2009, https://www.wired.com/2009/09/mf-deadhand/.

[49] Some accounts by former Soviet officials state that the dead hand system was investigated and possibly even developed, but never deployed operationally. See Interview with Gen.-Col. (Ret.) Andrian A. Danilevich, March 5, 1990, http://nsarchive.gwu.edu/nukevault/ebb285/vol%20iI%20Danilevich.pdf, 62-63; and Interview with Viktor M. Surikov, September 11, 1993, http://nsarchive.gwu.edu/nukevault/ebb285/vol%20II%20Surikov.PDF, 134-135. It is unclear, though, whether this refers in reference or not to a fully automatic system. Multiple sources confirm the system was active, although the degree of automation is ambiguous in their accounts. See Interview with Vitalii Leonidovich Kataev, 100-101; and Interview with Gen.-Col. (Ret.) Varfolomei Vladimirovich Korobushin, 207.

[50] Thompson, "Inside the Apocalyptic Soviet Doomsday Machine"; Interview with Vitalii Leonidovich Kataev; and Interview with Gen.-Col. (Ret.) Varfolomei Vladimirovich Korobushin.

[51] Ibid.

If the Soviets did indeed decide to develop and deploy a dead hand system, they did so without telling their American counterparts. This would appear, on the face of it, counterproductive. If the intent of a dead hand system was to enhance deterrence by ensuring an adversary that retaliation was certain, then keeping it a secret would appear to undermine the whole point of the system. According to reports from former Soviet officers, however, the point of Perimeter was not to change the decision-making calculus of U.S. leaders, but rather that of Soviet leadership themselves.[54] A dead hand system was intended to take the pressure off of Soviet leaders to "use or lose" their nuclear weapons in the event of ambiguous warning of a U.S. surprise attack. The logic of the Soviet approach illustrates how differently countries may view the role of automation in nuclear command and control. These differences in perspective, in turn, may lead a nuclear-armed state to misestimate or misunderstand the risks an adversary is willing to run in order to fortify their nuclear deterrent, thereby increasing chances of accidental or inadvertent escalation.

There are other examples of automation in NC2 systems. Bruce Blair has discussed how automation was used in the United States and Soviet nuclear enterprise. For example, Moscow used an "automated broadcast network"[55] to deliver battle orders in the case of a crisis or a first strike. In the United States, the Strategic Air Command Control System came online in 1968 to transmit Emergency Action Messages to U.S. forces in the field in case of a crisis. While the system had to be operated by a person, once activated and initiated, the system offered the United States the ability to transmit the message even if, subsequent to activation, U.S. command centers were destroyed.[56]

Russian missiles that were de-targeted following the end of the Cold War have been reportedly programmed to automatically revert to their wartime targets in the United States if launched without a flight plan. In the event of a deliberate decision by Russian leadership to launch, automation cut the time needed to re-target and launch all of their missiles to 10 seconds. Similarly, while U.S. missiles were reportedly set to targets in the middle of the ocean during peacetime, the entire U.S. missile arsenal could allegedly be retargeted in 10 seconds.[57] The automation-enabled ability to rapidly retarget missiles undoubtedly was a factor in Russian and American leadership being comfortable with de-targeting their missiles. Even so, if a Russian missile set to automatically revert to wartime targets were launched accidentally or without authorization, it could spark nuclear war. Dead hand systems or rapid retargeting of a nation's entire missile arsenal could thereby exacerbate the consequences of an accident or unauthorized use by making it easier for such events to automatically lead to catastrophe.

---

[52] Interview with Vitalii Leonidovich Kataev; and Interview with Gen.-Col. (Ret.) Varfolomei Vladimirovich Korobushin.
[53] Thompson, "Inside the Apocalyptic Soviet Doomsday Machine."
[54] Anthony M. Barrett, "False Alarms, True Dangers? Current and Future Risks of Inadvertent U.S.-Russian Nuclear War" (RAND Corporation, 2016), 10, http://www.rand.org/pubs/perspectives/PE191.html.
[55] Bruce G. Blair, *The Logic of Accidental Nuclear War* (Washington, D.C.: Brookings Institution, 1993), 129
[56] "Strategic Automated Command Control System-Data Transmission Subsystem (SACCS-DTS) Network Security Program," Air Force Instruction 33-107, Vol. 3 (June 1998), http://www.dtic.mil/dtic/tr/fulltext/u2/a405155.pdf; and "Information Technology: Federal Agencies Need to Address Aging Legacy Systems" (U.S. Government Accountability Office, May 2016), http://www.gao.gov/assets/680/677436.pdf.
[57] Bruce G. Blair, "Where Would All the Missiles Go?" *The Washington Post*, October 15, 1996.

In addition, automation of a state's nuclear command-and-control systems could be used to enhance deterrence by effectively tying one's hands, for instance, by communicating that any attack on a nation's homeland defense systems would trigger nuclear escalation. That is, a nuclear-armed state could view an explicit threat of automated retaliation as useful for escalation management. Autonomous systems – in particular, the expanded automation of a state's NC2 apparatus – could be used to increase uncertainty on the part of all involved in a conflict as to what it would take to trigger nuclear launch. A nuclear-armed state that is sufficiently risk tolerant and is confronted by a conventionally superior adversary may use this uncertainty to limit the scale or scope (i.e. geographic or targeting) of an attack on its interests.

Credibly communicating such a threat to an adversary might be challenging, however. Given that automation resides in software, its effects can be difficult to demonstrate prior to crisis or conflict.[58] If a state were to use automation to tie its hands but could not show that it had done so, it would be like "tearing out the steering wheel" in a game of chicken but being unable to throw it out the window.[59] The net effect of automation in this instance would be to reduce flexibility and increase crisis instability.

Opportunity for Improved Reliability?

These challenges with existing automation of nuclear command and control illustrate the way that automation can become a double-edged sword. Shortening the time and steps needed to launch nuclear weapons can help buy more time for decision-makers to weigh ambiguous information and make an informed judgment before acting. On the other hand, in the event of accidents or misuse, there may be fewer steps and consequently fewer safeguards in place.

A critical question is thus how militaries will employ advances in AI to influence their early warning and NC2 systems. There may be many places where militaries could employ new forms of autonomous systems to bolster the reliability and effectiveness of early warning and NC2. Human-machine teaming could help offset automation bias and thus enable the use of more autonomous systems. More advanced automation in nuclear early warning systems could allow greater situational awareness, reducing the risk of false alarms. It could also play a valuable role in helping human decision-makers process large amounts of information quickly. In this regard, automated data processing may play a critical role in helping human nuclear early warning operators to identify threats – and false cues – in an increasingly data-saturated and complex strategic environment. Increased automation in NC2 could also help to reduce the risk of accidents or unauthorized use. And an expanded role for automation in communications could help ensure that command-and-control signals reach their targets quickly and uncorrupted in highly contested electromagnetic environments.

---

[58] Michael C. Horowitz, "Artificial Intelligence, War, and Crisis Stability," *Journal of Strategic Studies*. Forthcoming.
[59] Herman Kahn, "On Escalation: Metaphors and Scenarios," (New York: Prager, 1965), 11.

Automation could also be used to enhance defenses – physical or cyber – against attacks on nuclear early warning, command-and-control, delivery, and support systems, thereby enhancing deterrence and fortifying stability. It could also be used to bolster the resilience of vulnerable NC2 networks. For instance, long-endurance uninhabited aircraft that act as pseudo-satellites ("pseudo-lites") to create an airborne communications network could increase NC2 resilience by providing additional redundant communications pathways in the event of satellite disruption. Automation could even enable autonomously self-healing networks – in physical or cyberspace – in response to jamming or kinetic attacks against command-and-control nodes, thereby sustaining situational awareness and command and control and enhancing deterrence.

Many of these ways that autonomous systems could increase the resiliency and accuracy of NC2 are speculative, however. Existing automation, as the Petrov incident shows, already creates the risk of automation bias. Knowledge of this will probably make most nuclear-armed states unlikely to further automate the early warning or command-and-control processes, with two exceptions: first, in situations where human-machine teaming might be further integrated to mitigate potential false alarms; second, in situations where a state fears for its secure second strike, and believes that further automation would reinforce deterrence of a potential aggressor. It is also possible, though less likely, that more automation could occur via a highly risk-tolerant nuclear-armed state that believes automated NC2 protocols would improve its ability to manage escalation.

Strategic Decision Support Systems

Strategic decision support systems could also affect nuclear stability by influencing how policymakers perceive and react to nuclear or strategic non-nuclear threats. States have long relied on computational methods to better understand threat environments and design solutions to emerging or imminent national security problems. During the Cold War, for instance, the Kremlin tasked the KBG with developing a computer program named "VRYAN" (the Soviet acronym for "Surprise Nuclear Missile Attack") that would track the U.S.-USSR correlation of forces and notify Soviet leaders when a preemptive nuclear strike would be required to prevent the United States from achieving decisive military superiority.[60]

VRYAN's role in providing strategic warning was tested during the Soviet "war scare." As early as the late 1970s, Soviet leaders were increasingly concerned that the United States had abandoned détente and had instead committed itself to achieving decisive military superiority. These fears climaxed in 1983 during NATO's annual command post exercise – codename "Able Archer 83" – with Moscow allegedly placing forces on higher readiness out of fear that the exercise was in fact the start of U.S. nuclear preemption.[61]

VRYAN's assessments fueled the Soviet leaders' concerns, however. The data used to feed the system was often prejudiced, as intelligence officers tended to submit information that conformed to the leadership's view that the United States was pursuing first-strike superiority.[62] VRYAN's assessments therefore reinforced Soviet leaders' fears about the United States, driving a feedback loop.[63] This loop may have been exacerbated by Soviet leaders' predominately engineering backgrounds, which may have predisposed them toward viewing the program's quantitative analysis as more credible than alternatives – a precursor to automation bias.[64] This feedback loop amplified and intensified those perceived threats, rather than providing Soviet leaders with a clearer understanding of U.S. intentions.

---

[60] President's Foreign Intelligence Advisory Board (PFIAB), "The Soviet 'War Scare,'" February 15, 1990, vi, 24 et seq, https://nsarchive2.gwu.edu/nukevault/ebb533-The-Able-Archer-War-Scare-Declassified-PFIAB-Report-Released/.
[61] PFIAB, "The Soviet 'War Scare,'" 7-8.

States' reliance on computational models today may be growing due to the AI revolution. In 2014, Russia erected the National Defense Control Center (NDCC) in Moscow. One of the NDCC's primary functions is information fusion in support of conventional and nuclear operations.[65] The Russian government is simultaneously investing heavily in AI, in part to better analyze the large quantities of data being delivered to NDCC and other agencies.[66] As in the 1980s, these investments are occurring at a time when Russian leaders see their nation as increasingly insecure due to the U.S. pursuit of military advantage, and the Russian military is more seriously evaluating the strategic merits of preemptive strike operations.[67]

China is similarly investing in AI-enabled decision support systems. As Lora Saalman writes, Chinese officials fear that the People's Liberation Army (PLA) would be unable to detect and counter a low-signature, prompt "bolt-from-the-blue" attack on its nuclear forces.[68] This fear reflects a combination of the perceived inadequacy of Chinese early warning systems and advances in U.S. prompt strike capabilities. The threat of a successful disarming attack on Chinese nuclear forces has led Chinese officials to prioritize avoiding "false negatives" over "false positives."[69] That is, whereas U.S. officials are concerned firstly by the potential for a "false positive" – whereby early warning systems show incorrectly that an attack is underway – their Chinese counterparts are more concerned by the possibility that early warning systems will show that an attack is *not* underway, when in fact it is.

China is investing heavily in AI-enabled decision support systems in part to help avoid false negatives by accelerating troops' ability to identify and respond to a disarming attack.[70] Officials' strong public emphasis on AI-supported decision-making as a potentially decisive innovation suggests that they may be prone to automation bias.[71] Chinese military-theoretical writing on information dominance and the notion of victory through scientific planning, each of which will rely on AI, according to researchers in China, also may make Chinese officials potentially susceptible to automation bias.[72]

---

[62] PFIAB, 81-82.
[63] Ibid.
[64] PFIAB, "The Soviet 'War Scare,'" 46.
[65] Tyler Rogoway, "Look Inside Putin's Massive New Military Command And Control Center," FoxtrotAlpha.com, November 19, 2015, https://foxtrotalpha.jalopnik.com/look-inside-putins-massive-new-military-command-and-con-1743399678.
[66] Samuel Bendett, "Here's How the Russian Military Is Organizing to Develop AI," Defense One, July 20, 2018, https://www.defenseone.com/ideas/2018/07/russian-militarys-ai-development-roadmap/149900/.
[67] Alexander Velez-Green, , "Russian Strategists Debate Preemption as Defense Against NATO Surprise Attack," Russia Matters, Belfer Center for Science and International Affairs, March 2018, https://www.russiamatters.org/analysis/russian-strategists-debate-preemption-defense-against-nato-surprise-attack.
[68] Lora Saalman, "Fear of false negatives: AI and China's nuclear posture," *Bulletin of the Atomic Scientists*, April 24, 2018, https://thebulletin.org/2018/04/fear-of-false-negatives-ai-and-chinas-nuclear-posture/.
[69] Saalman, "Fear of false negatives."
[70] Lora Saalman, "Lora Saalman on How Artificial Intelligence Will Impact China's Nuclear Strategy," *The Diplomat*, November 7, 2018, https://thediplomat.com/2018/11/lora-saalman-on-how-artificial-intelligence-will-impact-chinas-nuclear-strategy/.

Decision support systems are not inherently destabilizing. However, if states comes to over rely on these systems for strategic decision-making, this could undermine nuclear stability. This risk may grow, especially if decision-makers and their advisors believe that AI could serve as a panacea for the myriad informational problems (e.g. incomplete data or inadequate analysis) that have stymied their efforts at national defense over the years.

Moreover, AI-based decision-support systems may fail to deliver accurate information to decisionmakers precisely at the moment they are needed: in a crisis. Automated decision-support systems, whether rule-based or using machine learning, are only as good as the data they rely on. Building an automated decision-support tool to provide early warning of a preemptive nuclear attack is an inherently challenging problem because there is zero actual data of what would constitute reliable indicators of an imminent preemptive nuclear attack . This is most acutely challenging when trying to warn against a "bolt-from-the-blue." Intelligence services can track indicators of large-scale military mobilization. But these indicators cannot provide insight into the minds of senior decision-makers, who may not have yet made a decision whether to attack. Indeed, a nation readying its nuclear forces to launch a preemptive attack might appear similar to one preparing itself to "launch under attack" in response to what it perceives as indications that its adversary is preparing a nuclear first-strike.

Automation could be valuable in allowing intelligence agencies to scan large swaths of data quickly for anomalous behavior at a scale and speed that would not be possible with human analysts. This could be done through rule-based indicators where intelligence services set up the equivalent of automated alerts to warn when certain indicators are tripped. This signature-based approach is similar to how malware detection works today, where automation looks for known signatures of malicious software. Newer approaches using unsupervised machine learning can even assist in identifying anomalous activity when signatures are not yet known. These tools could be valuable in increasing the ability of intelligence services to track the digital footprint of military forces.

However, compiling AI-based indicators into an assessment of the likelihood of a preemptive attack would be extremely difficult, as the 1983 Stanislav Petrov incident highlights. Humans can rely on a multitude of contextual factors to help interpret indicators and assess an adversary's intent. Any rule-based system that attempted to make an assessment of the likelihood of an attack based on pre-specified indicators would be limited by the fact that human analysts who write the rules may not themselves know precisely what the actual indicators would be of a preemptive nuclear attack. Machine learning-based systems would similarly lack sufficient data to learn the signatures of a preemptive attack and could, at best, only indicate that some behavior was outside the norm.

---

[71] Elsa Kania, "Battlefield Singularity: Artificial Intelligence, Military Revolution, and China's Future Military Power," Center for a New American Security, November 2017, https://s3.amazonaws.com/files.cnas.org/documents/Battlefield-Singularity-November-2017.pdf?mtime=20171129235805.
[72] Burgess Laird, "War Control," Center for a New American Security, April 2017, https://s3.amazonaws.com/files.cnas.org/documents/CNASReport-ChineseDescalation-Final.pdf.

Even simple alert systems can be problematic if the manner in which they convey information is overconfident about the interpretation of that data and encourages automation bias, such as the Soviet early warning system communicating to Petrov "launch" and "missile strike."[73] Automated systems that more directly conveyed the information actually measured (in that instance, "flash") would decrease the risk of automation bias, by being more transparent to the human user. This tradeoff comes at a cost, however, as the human must take an additional step to interpret the data.

Governments thus face tradeoffs not only in whether or not to use automated decision-support tools, but in how that information is conveyed to human leaders. Automated decision support tools could be stabilizing if they help decision-makers gain better insights adversary's operations. This could help reassure leaders that an adversary is not planning an attack and could help make surprise attacks less feasible, reducing the incentives for preemption. On the other hand, false positives and automation bias could cause leaders to overreact to innocuous or ambiguous information, increasing instability. A major factor in how leaders calibrate such systems is likely to be their risk tolerance for false positives vs. false negatives. The more secure a country's second strike capabilities, the less likely it may be to take excessive risk with automating command and control, because the consequences of a false negative would be relatively lower. A country confident in its ability to retaliate in response to a first strike should be, on average, more likely to calibrate in ways that do not over rely on autonomous systems.

These risks of using automated decision-support systems are compounded by the fact that leaders won't have sufficient data about the systems' performance in a crisis to calibrate their degree of trust in it. Even worse, the system may perform adequately in peacetime, causing leaders to be lulled into a false sense of security about the system's reliability. Peacetime accuracy may cause leaders to place excessive faith in the systems' abilities to accurately identify and recommend responses to emergent threats. In this case, over time, leaders could become lulled into a sense of security about the efficacy of the system, even though they would have little actual data to support its value in warning of preemptive nuclear attack. Similar human-machine interaction failures have occurred in other settings where seemingly flawless performance leads humans to overtrust in automation, as in several fatal crashes involve Tesla autopilots. If a state does come to over rely on AI-enabled decision support systems for strategic decision-making, then it may fall subject to many of the limitations demonstrated in the VRYAN case. For instance, biased instructions for data collection to feed AI-enabled decision support systems may drive feedback loops that reinforce preexisting fears and amplify international tensions, potentially to the point of nuclear escalation. The potential for such loops may be increased if leaders believe that AI cannot be biased, and so take less care to remove their own biases from the design and use of the systems.

---

[73] Aksenov, "Stanislav Petrov."

*Nuclear Launch Platforms and Delivery Vehicles*

The second potential area where autonomous systems could influence nuclear stability is through their use as nuclear launch platforms and delivery vehicles. "Nuclear launch platforms," such as submarines, aircraft, missile launch facilities and associated control centers, or transporter erector launchers (TELs), are the systems that launch nuclear delivery vehicles. These platforms have historically been inhabited. The "nuclear delivery vehicle" is the bomb, missile, torpedo, or other mechanism that carries the nuclear warhead from launch to target. Nuclear delivery vehicles are not intended to be recoverable.[74]

Nuclear-armed states have historically limited the role of automation in nuclear launch platforms so that humans retain positive control over nuclear targeting and strike initiation. Yet this might change in the coming decades. The United States is building a new bomber, the B-21 Raider, which will be nuclear-capable and reportedly will be "optionally manned."[75] While the United States has not specified whether the B-21 bomber would ever be operated remotely while carrying nuclear weapons, statements by Air Force officials suggest such a move would be very unlikely. The Air Force's 2013 report, *Remotely Piloted Aircraft (RPA) Vector*, briefly mentions the issue, stating "nuclear strike, may not be technically feasible unless safeguards are developed and even then may not be considered for [unmanned aircraft systems] operations."[76] Informally, U.S. Air Force general officers have been more vocal about their discomfort with the concept of uninhabited vehicles armed with nuclear weapons. In 2016, General Robin Rand, head of Air Force Global Strike Command, stated: "We're planning on [the B-21] being manned. … I like the man in the loop … very much, particularly as we do the dual-capable mission with nuclear weapons."[77]

Other nations, however, may calculate the costs and benefits of using uninhabited nuclear launch platforms differently than the United States. For instance, states might choose to arm rapidly-proliferating UAVs with nuclear weapons if they saw benefits to doing so.[78] For example, in 2012, reflecting Russia's investments to overcome U.S. military superiority, a Russian Air Force lieutenant general stated Russia could field an uninhabited nuclear bomber in the 2040s.[79] One report suggests that North Korea may be considering using drones to disperse radioactive agents against the South Korean population in the case of a war.[80] Russia has also reportedly developed a seabed launcher for the Status-6 nuclear delivery vehicle, which is discussed further below.[81]

---

[74] That does not, however, mean that once a delivery vehicle is launched, then a nuclear attack is underway. If a delivery vehicle can loiter prior to receiving strike (vice launch) authorization, then its launch from a platform may not necessarily signal that an attack – nuclear or otherwise – is imminent. Christopher Ford, "Stability Engagement With Nuclear "Third Parties": Regional Risk Reduction Diplomacy," speech at "2019 Deterrence and Assurance Workshop and Conference," University of Nebraska, Omaha, NE, March 8, 2019, https://www.state.gov/t/isn/rls/rm/2019/290280.htm.

[75] Robert M. Gates, "Statement on Department Budget and Efficiencies" (U.S. Department of Defense, January 6, 2011), http://archive.defense.gov/Speeches/Speech.aspx?SpeechID=1527; and Dave Majumdar, "USAF leader confirms manned decision for new bomber," FlightGlobal.com, April 23, 2013, https://www.flightglobal.com/news/articles/usaf-leader-confirms-manned-decision-for-new-bomber-385037/.

[76] Headquarters U.S. Air Force, "RPA Vector: Vision and Enabling Concepts, 2013-2038" (February 17, 2014), 54, http://www.globalsecurity.org/military/library/policy/usaf/usaf-rpa-vector_vision-enabling-concepts_2013-2038.pdf.

[77] Hope Hodge Seck, "Air Force Wants to Keep 'Man in the Loop' with B-21 Raider," *Defencetech.org*, September 19 2016, https://www.defensetech.org/2016/2009/2019/air-force-wants-to-keep-man-in-the-loop-with-b-2021-raider/.

[78] On the general proliferation of UAVs, see Michael C Horowitz, Sarah E Kreps, and Matthew Fuhrmann, "Separating Fact from Fiction in the Debate over Drone Proliferation," *International Security* 41, no. 2 (2016): 7-42; Matthew Fuhrmann and Michael C Horowitz, "Droning On: Explaining the Proliferation of Unmanned Aerial Vehicles," *International Organization* 71, no. 2 (2017): 397-418..

Many nuclear delivery vehicles already incorporate automation into certain aspects of their operations. For instance, once an ICBM or submarine-launched ballistic missile (SLBM) is launched, the system relies on automation to maintain its flight trajectory and navigate by inertial or astro-inertial inputs to its assigned targets.[82] Open sources indicate, however, that some states are considering expanding the role of automation in their nuclear delivery vehicles.

For example, Russia is reportedly developing a "new intercontinental, nuclear-armed, nuclear-powered, undersea autonomous torpedo."[83] On March 1, 2018, Russian President Vladimir Putin confirmed the existence of an "unmanned underwater vehicle…that would carry massive nuclear ordnance."[84] He characterized the weapon, known as "Status-6" or "Poseidon," as a response to U.S. investments in missile defenses, which Russian leaders fear could be used alongside long-range non-nuclear precision strike weapons to neutralize Russia's nuclear deterrent.[85] Putin went on to say that Status-6 is designed for use against aircraft carrier groups or coastal targets, and that it would rely on speed, depth, maneuverability, and quietness to reach its target.[86] The Russian government has said little about the role of AI in Status-6, but a Russian source suggests that the Status-6 might be able to use AI to evade enemy anti-submarine warfare (ASW) forces on the way to its target.[87]

A Boon to Escalation Management?

---

[79] "Russia Could Deploy Unmanned Bomber After 2040 - Air Force," RIA Novosti, February 8, 2012, http://www.globalsecurity.org/wmd/library/news/russia/2012/russia-120802-rianovosti01.htm.
[80] Kyle Mizokami, "Experts: North Korea May Be Developing a Dirty Bomb Drone," *Popular Mechanics*, December 28, 2016, http://www.popularmechanics.com/military/weapons/a24525/north-korea-dirty-bomb-drone/.
[81] H.I. Sutton, "Poseidon Torpedo," Covert Shores, February 22, 2019, http://www.hisutton.com/Poseidon_Torpedo.html.
[82] Lieber and Press, "The New Era of Counterforce: Technological Change and the Future of Nuclear Deterrence; Glaser and Fetter, "Should the United States Reject Mad? Damage Limitation and Us Nuclear Strategy toward China." "Ballistic Missile Basics," Federation of American Scientists, https://fas.org/nuke/intro/missile/basics.htm; Jane Gibson and Kenneth G. Kemmerly, "Intercontinental Ballistic Missiles," in Air Command and Staff College, "AU-18 Space Primer" (Maxwell, AL: Air University Press, September 2009), 235-248, https://media.defense.gov/2017/Mar/15/2001717230/-1/-1/0/AU-18.PDF.
[83] U.S. Department of Defense, "Nuclear Posture Review 2018," 8-9.
[84] President of Russia, "Presidential Address to the Federal Assembly," March 1, 2018, http://en.kremlin.ru/events/president/news/56957.
[85] Ibid.
[86] Ibid. H.I. Sutton disputes the claim that Status-6 would be particularly quiet. See H.I. Sutton, "Countering Russian Poseidon Torpedo," Covert Shores, August 15, 2018, http://www.hisutton.com/Countering_Russian_Poseidon_Torpedo.html.
[87] Vladimir Tuchkov, "'Status-6': Retaliatory weapon that drove the Pentagon into a stupor," Free Press, December 11, 2016, https://svpressa.ru/war21/article/162378/. H.I. Sutton disputes this reporting, as well, writing, "[Status-6's] operating modes and route planning will likely be simple (read *reliable*) and relatively direct [emphasis original]." See Sutton, "Countering Russian Poseidon Torpedo."

Uninhabited nuclear launch platforms may be seen to offer some strategic benefits to nuclear-armed states. Nuclear-armed UAVs, for instance, could be kept aloft for far longer than is possible with human pilots, decreasing fear of a disarming first strike. B-2 bomber pilots, for instance, have flown a maximum of 40-hour missions.[88] By contrast, refuelable UAVs could stay aloft for several days, limited only by engine lubricants and other reliability factors. Uninhabited aircraft have already conducted 80-hour flights.[89] The maximum endurance record for a refuelable aircraft is 64 *days*.[90]

The ability to keep nuclear bombers in the air for longer periods of time might offer policymakers new tools for managing escalation. Long-endurance nuclear-armed UAVs could provide policymakers with additional options for nuclear signaling, since they could be kept on-station longer than would otherwise be possible. Likewise, if they are sufficiently survivable against adversary countermeasures, nuclear-armed UAVs might improve a state's ability to deliver nuclear weapons in a timely manner since they could be kept aloft closer to potential targets longer than their manned counterparts. For some less powerful nuclear-armed states, UAVs may also be seen as a lower-cost, longer-range alternative to human-inhabited nuclear bombers. Lower-cost systems are unlikely to be as survivable as their more expensive counterparts, however, thus limiting their utility.

Nuclear delivery vehicles that leverage AI for certain functions may also be seen to provide strategic benefits. For instance, the Status-6's notional AI-enabled counter-ASW capabilities may help to improve Russian leaders' confidence in their secure second strike regardless of advances in U.S. missile defenses by convincing them that their nuclear-armed torpedoes will always be able to reach their targets. This might constitute an improvement to U.S.-Russian nuclear stability.[91] But any such reassurance will be limited by the fact that, while torpedoes may pose a threat to coastal targets, they cannot strike inland strategic targets, such as enemy leadership redoubts, command centers, strategic forces, critical infrastructure, or population centers.[92] As a result, even if automation does improve Status-6 survivability, it would constitute only a marginal improvement to the overall viability of Russia's nuclear deterrent, since from Moscow's perspective, U.S. missile defenses and strike capabilities could still prevent it from using missiles to hold the full range of necessary targets at risk.[93]

---

[88] "B-2 bombers lead 'Shock and Awe,'" *Fox News,* March 26, 2003, http://www.foxnews.com/story/2003/03/26/b-2-bombers-lead-shock-and-awe.html.
[89] Graham Warwick, "Aurora claims endurance record for Orion UAS," *Aviation Week,* January 22, 2015, http://aviationweek.com/defense/aurora-claims-endurance-record-orion-uas.
[90] The record was set in 1958 in a Cessna 172. "Endurance test, circa 1958," *AOPA,org,* March 1, 2008, https://www.aopa.org/news-and-media/all-news/2008/march/01/endurance-test-circa-1958.
[91] Glenn Snyder's "stability-instability paradox" suggests that it might also cause the Kremlin to become more risk-tolerant, thereby destabilizing the U.S.-Russian strategic relationship. See Glenn Snyder, "The Balance of Power and the Balance of Terror," in *The Balance of Power*, ed. Paul Seabury (San Francisco, CA: Chandler, 1965).
[92] For more on nuclear targeting, see Matthew G. McKinzie, Thomas B. Cochran, Robert S. Norris, and William M. Arkin, "The U.S. Nuclear War Plan: A Time for Change" (Natural Resources Defense Council, June 2001), https://www.nrdc.org/sites/default/files/us-nuclear-war-plan-report.pdf.
[93] Importantly, this limitation may not feature as strongly in other nuclear dyads. If a country assesses that enough of its adversary's strategic targets are coastally located, or could be decisively affected by radiological fallout or other effects traveling inland from a coastal detonation, then it may rightly view nuclear-armed torpedoes as providing a greater-than-marginal improvement to its nuclear deterrent.

The Status-6 would be similarly constrained as a limited nuclear option for use against land targets, thereby further minimizing its likely impact on nuclear stability. If the weapon actually carries a "massive nuclear ordnance," as Putin said, then it is unlikely to be used against a land target during a limited nuclear war. That is, so long as the nuclear first-user's objective was to impel a de-escalation of hostilities – a logic evident in Russian thinking[94] – then using a high-yield area-devastation weapon would likely be seen as counterproductive, since it is more likely to invite reciprocal escalation. If the Status-6 warhead's yield were lower, Russian leaders might view the delivery vehicle's AI-enabled counter-ASW capabilities as helping to ensure their ability to execute limited nuclear strikes on the U.S. homeland. Even if true, however, the weapon would still only be able to reach coastal targets. Furthermore, if a Status-6 target was a population center, critical infrastructure, or a major military facility – or if the collateral effects of a Status-6 attack rose above some threshold – then the U.S. would be less likely to perceive the attack as being limited and might respond accordingly. Therefore, while the weapon might help to reassure Russian leaders of their ability hold the U.S. homeland at risk of limited nuclear strike, any confidence gained – and any consequent effects on nuclear stability – is likely to be marginal.

The Kremlin might view Status-6 as more significantly enhancing its ability to use limited nuclear force at sea. If Russian leaders believe that Status-6's notional AI-enabled counter-ASW capabilities improve their ability to hold naval targets at risk of nuclear strike, then they might be more inclined to consider or execute such a strike during a crisis or conflict. Yet, whether that would be stabilizing or not would be highly contingent on the circumstances. For instance, were Russian leaders to credibly communicate that inclination to their Western adversaries, fear of Russian nuclear use at sea might impel restraint on the latter's part, thereby reducing the likelihood of further escalation in a crisis or conflict. Alternatively, Washington and its allies might rededicate themselves to denying Russian leaders' any confidence in their ability to use nuclear force to impel war termination on terms favorable to them, for instance, by promising a like response should Russia cross the nuclear threshold. In that case, by providing Russian leaders with a new tool for nuclear brinksmanship, the Status-6 might contribute to a scenario where reciprocal nuclear escalation is suddenly more plausible.

Escalation Risks

Uninhabited nuclear launch platforms would pose challenges for maintaining nuclear surety, or positive human control, over nuclear weapons. Recoverable uninhabited platforms differ fundamentally from missiles because the former can be sent on patrol. With uninhabited launch platforms sent on patrol, policymakers would have effectively delegated the decision to start nuclear war to some combination of the system's remote communications links to human operators and onboard automation. There would be no human onboard to have physical control over the nuclear weapons, thereby degrading surety. While militaries could employ encrypted communications links and fail-safe measures such as automation directing the platform to return home in the event of a communications failure, these measures cannot always guarantee effective control. In 2011, for instance, a U.S. RQ-170 stealth drone crashed in Iran, landing sensitive technology in the hands of the Iranian government.[95] At worst, loss of control of an uninhabited nuclear launch platform due to technical malfunction or an unanticipated environmental interaction could set the stage for accidental escalation, setting nations on the path to nuclear escalation.[96] Short of that, it could lead to loss of nuclear weapons, loss of sensitive technologies, and a diplomatic crisis.

---

[94] Johnson, "Russia's Conventional Precision Strike Capabilities, Regional Crises, and Nuclear Thresholds."

Not only could a state potentially lose control of a system, the system would also be vulnerable to hacking. While digital systems are vulnerable to hacking regardless of whether there is a person onboard, uninhabited systems raise unique cyber security challenges. Without a person onboard who has physical access to the system and could disable it or employ hardware-level cutouts to critical systems, an adversary could potentially take control of an uninhabited vehicle. Intrusions could occur via communications links, but even "off network" systems are vulnerable to hacking.[97] An adversary could gain physical access to the system or implant malware via ground maintenance software.[98] Severing a vehicle's communications links and making it fully autonomous would reduce some avenues for intrusion, but not eliminate them entirely, and would come at the cost of losing the ability to recall the system.[99] While less likely, in the worst case, hacking uninhabited nuclear launch platforms raises the potential for catalytic escalation, whereby a third-party actor hacks a state's military system and uses it to attack another country, thereby triggering escalation between the targeted states.

Many of the same risks would be present in nuclear delivery vehicles that leveraged AI but relied on external communications or other inputs for mission execution. For instance, once Status-6 is deployed, if it is capable of receiving instructions from remote operators, then it may also be more vulnerable to technical malfunction, unanticipated environmental interactions, or third-party interference, each creating new risk of accidental or catalytic escalation.

A nuclear delivery vehicle that could autonomously select its target would introduce additional risks. Even if a human deliberately launched a nuclear strike, if the weapon hit the wrong target it could unintentionally escalate a conflict or even hit a target in a third country. There is currently no evidence to suggest that Status-6 is capable of autonomous target selection.

Status-6 could also threaten nuclear stability for reasons driven less by automation than by other aspects of its design or use. One reason is that its lengthy transit time could complicate war termination efforts. That is, assuming Status-6 maintains its reported top speed of 70 knots and takes the most direct route, it would take it over a day and a half to reach New York from the

---

[95] Bob Orr, "U.S. official: Iran does have our drone," *CBS News*, December 8, 2011, http://www.cbsnews.com/news/us-official-iran-does-have-our-drone/.
[96] Alexander Velez-Green, "The nuclear mission must stay manned," *Bulletin of the Atomic Scientists*, August 9, 2016, http://thebulletin.org/nuclear-mission-must-stay-manned9768.
[97] Nicolas Falliere, Liam O'Murchu, and Eric Chien, "W.32 Stuxnet Dossier," *Symantec Security Response*, February 2011, https://www.symantec.com/content/en/us/enterprise/media/security_response/whitepapers/w32_stuxnet_dossier.pdf.
[98] Jacquelyn Schneider, "Digitally-Enabled Warfare: The Capability-Vulnerability Paradox," *Center for a New American Security*, August 2016, https://www.cnas.org/publications/reports/digitally-enabled-warfare-the-capability-vulnerability-paradox.
[99] One option would be reversion to environmental signatures for communication (e.g. stigmergy). But may be compromised by enemy countermeasures or other environmental inputs.

Denmark Strait. That transit time increases to over two days if it is launched from the Barents Sea.[100] If the weapon is not recallable, then once it is launched, the Russian government's ability to negotiation war termination will be seriously limited by the fact that a nuclear strike is already en route to the U.S. homeland even as they are seeking off-ramps. This risk would be mitigated somewhat if the weapon were launched closer to American shores – and coastal defenses. But even if it were launched from near Bermuda, for example, it would still take over 10 hours to arrive – a not-insignificant amount of time in the midst of a nuclear war.[101] This risk could also be mitigated if Status-6 is recallable. But recallability – like any technical function – may be vulnerable to failure. That vulnerability is accentuated by the fact that radio communications are difficult underwater.[102] Furthermore, if the system is recallable, this might create new operational risks due to unanticipated environmental interactions or hacking.

Thus, the area of uninhabited nuclear delivery platforms and more highly-automated nuclear delivery vehicles seems like a critical one – especially given the potential for accidents and miscalculation described above. Even though the United States prioritizes nuclear surety, and is therefore unlikely to deploy uninhabited nuclear delivery platforms, other nations may not show similar reticence. The same can be said for nuclear delivery vehicles featuring higher levels of automation.

*Conventional Military Applications of Autonomous Systems*

Autonomous systems could also affect nuclear stability through their conventional military applications. Among major powers, the Third Offset strategy launched by the Department of Defense during the Obama administration highlights the way the United States military believes that autonomous systems could influence the future of conventional warfare.[103] Translated Chinese military and industry documents, for example, demonstrate intense research and development by China regarding the deployment of autonomous combat systems in the air and in other domains.[104] The Russian Federation is similarly investing in military robotics in order to strengthen its conventional deterrent against NATO or other foes. Moreover, states are likely to be more willing to accept increased autonomy in conventional systems than nuclear ones, allowing potentially greater opportunities for these applications.[105] Nevertheless, some conventional applications of automation could possibly affect nuclear stability in significant ways that are both direct and indirect.

---

[100] At 70 knots, Status-6 would need about 41 hours to travel 2,900 miles from the Denmark Strait to Lower New York Bay. At the same speed, it would need about 57 hours to travel 4,000 miles from Bear Island to Lower New York Bay. See Sutton, "Poseidon Torpedo."

[101] At 70 knots, Status-6 would need about 10.5 hours to travel 750 miles from Bermuda to Lower New York Bay. See Sutton, "Poseidon Torpedo."

[102] Emma M. O'Shaughnessy, "Characterising the Relative Permittivity and Conductivity of Seawater for Electromagnetic Communications in the Radio Band - Summary Report 2012," School of Engineering and Information Technology, UNSW@ADFA, http://ojs.unsw.adfa.edu.au/index.php/juer/article/download/621/393.

[103] Robert O. Work, "National Defense University Convocation: As Prepared for Delivery by Deputy Secretary of Defense Bob Work, National Defense University, Washington, Dc, Tuesday, August 05, 2014," (http://www.defense.gov/speeches/speech.aspx?speechid=1873: U.S. Department of Defense, 2014).

[104] Graham Webster et al, "China's Plan to 'Lead' in Ai: Purpose, Prospects, and Problems," *New America Foundation Cybersecurity Initiative*, August 1, 2017 2017, https://www.newamerica.org/cybersecurity-initiative/blog/chinas-plan-lead-ai-purpose-prospects-and-problems/.. The full text of China's AI development plan is available at https://www.newamerica.org/documents/1959/translation-fulltext-8.1.17.pdf

Counterforce Operations

Direct uses of autonomous systems for the nuclear mission intersects with ongoing debates about how technological change is influencing nuclear survivability. Daryl Press and Keir Lieber argue that changes in accuracy and transparency are decreasing arsenal survivability in ways that are increasing the potential for counterforce strikes.[106] They state that robotics and AI could allow for real time tracking of adversary nuclear arsenals and rapid, accurate strikes in ways that make counterforce operations more plausible.

However, there are several reasons to think that while specific applications of AI might improve the ability of countries to find and fix adversary nuclear arsenals in some contexts, this is unlikely to significantly change the potential for counterforce strikes. Their argument relies on two key assumptions. On accuracy, they presume that a nuclear war will occur in a world where countries have full access to their targeting sensors and data – and where those sensors and any resulting data, in turn, are not subject to adversary countermeasures. This presumes more of a bolt from the blue event than an escalation of a conventional conflict. On transparency, they presume that 24 minute-interval satellite passes over roadways, in combination with longer-range aerial surveillance from UAVs and other assets, would provide enough information to generate real-time tracking against mobile Transporter Erector Launcher (TEL) units employing deception and concealment. In principle, if this capability were robust, it could decrease states' confidence in the survivability of their mobile ICBMs, undermining deterrence.[107] Thus, a conventional application of AI, in the ISR collection and processing arena, could have strategic consequences.[108]

Key hurdles, however, are achieving the scale and robustness needed to find, fix, and track not just one target but all of a country's mobile nuclear systems, as well as the timeline for doing so. In order to be relevant, the information must be collected and processed with high fidelity, at scale, and in a timely enough fashion to shape nuclear strike decisions. Machine learning algorithms, particularly those trained on huge datasets of past surveillance data, could potentially increase the speed of processing data on adversary nuclear locations. It is far from clear, however, that the application of AI would provide enough of an improvement in transparency to change how countries think about whether to use nuclear weapons. As Miller, Fontaine, and Velez-Green write, "[I]t is one thing to locate a system, for instance in the middle of the Atlantic Ocean or the Siberian forest. It is another thing to be able to deliver a sufficiently destructive and accurate weapon against the targeted system before it is able to fire or conceal itself."[109] The destructive power of nuclear weapons means that even one missed weapon could wipe out a city. Even a low probability that a single weapon might survive a first strike and be used in a retaliatory strike could be enough to deter the attempt. Nevertheless, perceptions matter, and if a state perceived its arsenal to be vulnerable, that may change state behavior in a crisis.

---

[105] Michael C. Horowitz, "Military Robotics, Autonomous Systems, and the Future of Military Effectiveness," in *The Sword's Other Edge: Tradeoffs in the Pursuit of Military Effectiveness*, ed. Dan Reiter (New York: Cambridge University Press, Forthcoming).
[106] Lieber and Press, "The New Era of Counterforce: Technological Change and the Future of Nuclear Deterrence." Also see Keir A. Lieber and Daryl G. Press, "The End of Mad? The Nuclear Dimension of U. S. Primacy," *International Security* 30, no. 4 (2006): 7-44.
[107] Austin Long and Brendan Rittenhouse Green, "Stalking the Secure Second Strike: Intelligence, Counterforce, and Nuclear Strategy," *Journal of Strategic Studies*, 38 no. 1-2 (2015), 38-73; and Paul Bracken, "The Cyber Threat to Nuclear Stability," *Orbis*, 60 no. 2 (2016). For an application to AI, see Geist and Lohn, "How Might Artificial Intelligence Affect the Risk of Nuclear War" and Technology for Global Stability, "Artificial Intelligence and the Military: Forever Altering Strategic Stability,"
[108] Geist and Lohn, "How Might Artificial Intelligence Affect the Risk of Nuclear War."

Some argue that networks of UUVs, USVs, and UAVs could dramatically improve states' abilities to hold strategic missile-armed nuclear-powered submarines (SSBNs) at risk, threatening the latter's deterrent credibility.[110] However, similarly the complexity of the ASW mission – the steps that must be taken for ASW forces to detect, classify, localize, and engage a single SSBN, much les an entire fleet of SSBNs – means these risks also are likely overstated. Generally, to hold an SSBN at risk, ASW forces must persistently maintain sensors where they believe that submarine is likely to operate, share sensor and other data in a timely manner to coordinate to maintain optimal coverage, cue and direct searches, and confidently classify an underwater contact as an SSBN. Once ASW forces localize an SSBN, they must maintain localization until an ASW weapons-carrying platform is guided to attack range of the SSBN and consummates an engagement. Depending on the situation, this process can take hours or even days.[111]

Claims that AI could generate a "transparent ocean"[112] or "selective ocean transparency"[113] likely overstate the ability of low-cost UUVs, USVs, and UAVs to conduct these steps. First, in order to be low-cost, uninhabited vehicles are generally of limited size, weight, and power (SWaP) capacity, at least relative to traditional attack submarines. These facts, combined with the inherent physics-based difficulties of sensing in the undersea environment,[114] mean that the sensors carried by any given low-cost UUV, USV, or UAV will be of limited detection range regardless of the phenomenologies they employ.[115] Given the sensors' limited range, ASW forces would need to deploy large numbers of sensor vehicles to seamlessly cover even small oceanic areas.[116] UUVs in particular are limited in endurance, because of their need to rely on air-independent power sources such as batteries or fuel cells. This means that additional UUVs would be needed to sustain a track on a submarine over time, as UUVs reach the end of the their operational endurance and need to return to base. (Ultra-long endurance UUV power solutions, such as thermal gliders that draw energy from ocean thermoclines, lack sufficient speed and power to maintain track on a submarine.[117]) While fleets of UUVs, USVs, and UAVs are likely to have cost-savings relative to traditional assets and will be valuable supplements in a "high-low mix" of ASW capabilities, the scale of assets needed to render even a portion of the ocean "transparent" would likely be enormous. Setting cost and practicality aside, ASW forces would also have to keep these sensors appropriately positioned to maintain high-confidence area surveillance and target tracking. This would require a level of multi-system control reliability and resilience not yet demonstrated.[118]

---

[109] James N. Miller, Jr., Richard Fontaine, and Alexander Velez-Green, "Averting the U.S.-Russia Warpath," *The National Interest*, March/April 2018, https://nationalinterest.org/feature/averting-the-us-russia-warpath-24604.
[110] Lieber and Press, "The New Era of Counterforce". For more discussion, see Gates, "Is the SSBN Deterrent Vulnerable to Autonomous Drones?"
[111] For more on ASW, see Owen R. Cote Jr., *The Third Battle: Innovation in The Navy's Silent Cold War Struggle With Soviet Submarines* (Newport, RI: US Naval War College Press, 2003).
[112] David Hambling, "The Inescapable Net: Unmanned Systems in Anti-Submarine Warfare," Briefing No. 1, Parliamentary Briefings on Trident Renewal (BASIC, March 2016).
[113] Sebastian Brixey-Williams, "Will the Atlantic become transparent?" Third Edition (British Pugwash, 2018).
[114] Rear Admiral Charles Richard, "The Mirage of Transparent Ocean," Navy Live, July 29, 2016, https://navylive.dodlive.mil/2016/07/29/the-mirage-of-a-transparent-ocean/.
[115] Jonathan Gates, "Is the SSBN Deterrent Vulnerable to Autonomous Drones?" *The RUSI Journal,* 161 no. 6 (2016), 29-33; and Ian Keddie, "Trident at Risk? UMV Technology vs Submarine Stealth," *CABLE Magazine* (2017).
[116] Rear Admiral John Gower, "Concerning SSBN Vulnerability" (BASIC, June 2016).

Fewer sensors would be required to monitor ocean chokepoints. But only Chinese and British SSBNs must pass through chokepoints to hold their primary targets at risk – and both countries have offset this risk. Although Chinese SSBNs would need to pass through chokepoints in the First Island Chain to cover the entire United States, China's land-based mobile ICBM force can cover targets at that range. Likewise, British SSBNs must pass through chokepoints to the north or south of Ireland to reach deep waters. But, as British Rear Admiral John Gower has written, monitoring those chokepoints would probably still require a high number of sensors.[119] And the costs of maintaining or cycling those systems would still be high. Furthermore – and crucially – any UUVs, USVs, or UAVs deployed in the chokepoints would be subject to countermeasures employed on an adaptive basis, including improved stealth, jamming, multi-phenomenology decoys and spoofing, evasive maneuvers, or outright destruction by SSBN protection forces.[120] Many of the same countermeasures could also be used against sensors operating in the open ocean.

Even if ASW forces were able to maintain optimal sensor coverage in the search area, they would then face problems of data transmission that automation is ill suited to solve. The UUVs, USVs, and UAVs sent to monitor the open ocean or a chokepoint must be able to share data – processed or raw – amongst themselves – directly or through command nodes – in a timely manner in order to maintain coverage, cue and direct searches, confidently classify a contact as an SSBN, and then support weapons employment against the SSBN. In order to be used effectively, any vehicle that attains a track on an SSBN would need to be able to transmit that data to another vehicle.[121]

While highly automated network management technologies may be able to enhance communications resilience between uninhabited – or uninhabited and inhabited – ASW platforms,[122] such data transmission will remain a vulnerability for any undersea communications.[123] It bears noting that the physics of undersea communications results in fairly short range communications paths at low data rates. Longer range communications paths and higher data rates forces reliance on surface or airborne communications relays that are vulnerable to jamming or other interference. Even temporary or partial delays in data transmission could undermine ASW forces' abilities to localize an SSBN – and given that the window of opportunity to localize a submarine may be very short, an inopportune communications delay or disruption may make the difference between ASW success and failure.[124]

---

[117] "AUV System Spec Sheet: Slocum thermal glider configuration," AUVAC: Autonomous Undersea Vehicle Application Center, https://auvac.org/configurations/view/51.
[118] Bradley Martin, Danielle C. Tarraf, Thomas C. Whitmore, Jacob DeWeese, Cedric Kenney, Jon Schmid, and Paul DeLuca, "Advancing Autonomous Systems: An Analysis of Current and Future Technology for Unmanned Maritime Vehicles" (RAND Corporation, 2019).
[119] Gower, "Concerning SSBN Vulnerability."
[120] Gates, "Is the SSBN Deterrent Vulnerable to Autonomous Drones?" 30; Gower, "Concerning SSBN Vulnerability"; Keddie, "Trident at Risk?"; and Bryan Clark, "The Emerging Era in Undersea Warfare" (Center for Strategic and Budgetary Assessments, January 2015), 15.
[121] Clark, "The Emerging Era in Undersea Warfare," 8, 10.

Finally, if we assume that ASW forces relying on UUVs, USVs, and UAVs were able to confidently classify and localize an SSBN in their search area, they would need to maintain localization long enough for an ASW weapons-carrying platform to close within attack range of the SSBN and successfully engage it. If a weapons-carrying platform is located close by, for instance, near a chokepoint through which a SSBN is transiting and where that SSBN's protection forces are unable provide effective coverage – again, an implausible scenario for reasons of both geography and nuclear force structure for all nuclear-armed states – this may be a relatively easy problem to solve. However, if the search is occurring in the open ocean, the sheer expanse of that area suggests that a weapons-carrying platform is unlikely to be within immediately-actionable proximity of the SSBN when confident classification is made.[125]

Automated protocols might reduce the time required to signal and dispatch an ASW weapons-carrying platform. But the weapons-carrying platform's transit time alone would leave a window for the SSBN's crew to conduct countermeasures—or for inherently dynamic underwater conditions to degrade the ability of the sensors in contact to maintain track. This time window could be reduced by arming uninhabited vehicles directly. Nevertheless, the problems of coordinating multiple vehicles, at scale, for an extended period of time, and robustly in a challenging communications environment amidst adversary countermeasures remains.

If ASW forces miss any of these steps, then they will be unable to detect, classify, localize, and engage the SSBN. To successfully prosecute a disarming first strike against a nation's entire SSBN fleet, an opposing nation's ASW forces would need to execute the entire kill chain for every one of those boats – and probably near-simultaneously to avoid triggering fleet-wide countermeasures (which would render subsequent ASW operations even more difficult) or strategic escalation.[126]

---

[122] Paul Scharre, "Unleash the Swarm: The Future of Warfare," War on the Rocks, March 4, 2015, https://warontherocks.com/2015/03/unleash-the-swarm-the-future-of-warfare/.

[123] On the importance of oceanic topography for undersea surveillance, see Owen R. Cote, Jr., "Invisible nuclear-armed submarines, or transparent oceans? Are ballistic missile submarines still the best deterrent for the United States?" *Bulletin of the Atomic Scientists*, 75 no. 1 (2019).

[124] Keddie, "Trident at Risk?"

[125] Contrary to Hambling's assertion, "A submarine whose location is exposed is highly vulnerable to instant attack." See Hambling, "The Inescapable Net," 1.

[126] "It is safe to assume the Chinese navy will implement CASD [continuous at sea deterrent] patrols in the foreseeable future: this suggests that the Chinese leadership has confidence that their considerable investment in an SSBN capability will be secure in the long term." See Keddie, "Trident at Risk?" Richard wrote to similar effect: "The SSBN Security Technology Program and SSN/SSGN Survivability Program identify emerging risk areas and map out proactive mitigation plans years before anything starts showing up in the mainstream media." See Richard,

There might also be other ways states could leverage increased automation, autonomy, and artificial intelligence for counter-nuclear operations as well. These could include enhanced missile defenses, improving the accuracy of conventional or nuclear counterforce strike options, or boosting the efficacy of offensive cyber-attacks against enemy nuclear command-and-control systems. Like the ASW argument, it is entirely possible that many of these concepts, while theoretically possible, are not realistic or feasible given the state of technology today or on the near-horizon. Even if these capabilities do not materialize, however, the current rapid pace of technological advancement may increase the risk of misperception. If a nuclear-armed state's leadership comes to believe that military automation has or may soon have the ability to undermine its nuclear deterrent – whether that is correct or not – it may take destabilizing hedging measures, such as heightening alert statuses or delegating nuclear launch authority.

Speed in Conventional War

In addition, autonomous systems could compress decision cycles in conventional warfare. As former U.S. Deputy Secretary of Defense Robert O. Work has said, algorithms will be able to make decisions faster than humans, propelling the pace of armed conflict to "machine speed."[127] States may react to these shifts in dangerous ways. For instance, if a nuclear-armed state fears that its adversary might launch a disarming attack on its nuclear forces before it can respond, then it might further automate nuclear launch processes, put its nuclear arsenals on heightened alert, or pre-delegate more authority to field commanders. It might also endorse a doctrine of preemption, so as to get ahead of an adversary disarming attack.[128] These moves, designed to deter, defeat, or preempt a disarming strike and give their country the ability to strike back in the face of conventional defeat, might lead one side to use nuclear weapons first. And, in the event of accidents or intentional misuse, automation could shorten the steps needed for an incident to escalate.

The potential for nuclear escalation in a conventional conflict with autonomous systems is compounded by the way that autonomous systems could enable adopters to fight faster than those operating non-autonomous systems do at present. Because of machines' superior reaction times, autonomous systems would have a speed-based edge in decision-making over human-controlled systems. This could have immediate tactical advantages. An autonomous plane might be more adept at identifying and avoiding air defense threats, for example, or better at predicting and defeating adversaries in an air-to-air engagement, making it more able to complete its mission. The cumulative effect of this faster decision-making could also translate into a faster tempo at the operational level of war. A military force that is heavily invested in AI could

---

"The Mirage of Transparent Ocean." For more on Russian nuclear modernization, see Hans M. Kristensen & Robert S. Norris, "Russian nuclear forces, 2018," *Bulletin of the Atomic Scientists*, 74 no. 3 (2018).

[127] Robert O. Work, "As Delivered by Deputy Secretary of Defense Bob Work, Washington, D.C, Oct. 4, 2016," Deputy Secretary of Defense Speech at Association of the United States Army Annual Convention, October 4 2016, https://www.defense.gov/News/Speeches/Speech-View/Article/974075/remarks-to-the-association-of-the-us-army-annual-convention/.. Also see Michael C. Horowitz, "The Future of War Is Fast Approaching in the Pacific: Are the U.S. Military Services Ready?," *War on the Rocks*, 2017, https://warontherocks.com/2017/2006/the-future-of-war-is-fast-approaching-in-the-pacific-are-the-u-s-military-services-ready/.

[128] Velez-Green, "The Unsettling View from Moscow"; and Velez-Green, "Russian Strategists Debate Preemption as Defense Against NATO Surprise Attack."

essentially enable faster operations by autonomous systems relative to remotely-piloted or inhabited systems. Some Chinese scholars have hypothesized that this trend could result in a "battlefield singularity," in which the pace of action on the battlefield eclipses the speed of human decision-making.[129]

Fear of losing quickly could create incentives for more rapid escalation to the nuclear level. To the extent that reasoned, sustained thinking makes countries more likely to back away from the nuclear brink, autonomous systems, and especially autonomous weapon systems, could undermine the security of time. The fear of losing – or falling behind the pace of battle and losing the ability to even identify when a loss is imminent – could lead countries to take drastic measures in response. If leaders perceived regime security or nuclear command-and-control were at risk, such time-sensitive pressure could incentivize preemptive escalation or, in the worse cases, nuclear first-use.[130]

It is also possible, however, that autonomous systems could help countries buy time in ways that make nuclear escalation less likely. The United States and its competitors already deploy expansive sensor networks, stretching from outer space to cyberspace. But the amount of data produced by these sensing arrays threatens to overwhelm human operators today. The emergence of highly-automated data processors could change that, allowing national decision-makers to make far better sense of the cluttered battlespace. In fact, the first use of "AI" technology at the U.S. Department of Defense was for automated information processing to help monitor full-motion video drone feeds through Project Maven.[131] Greater awareness and understanding of an adversary's actions could reduce the risk of miscalculation. Leaders would be able to replace uncertainty – and a fear of the worst – with near real-time information on an adversary's forces. Greater visibility could reassure leaders that a surprise attack was not underway, and the knowledge of this visibility would reduce incentives for a surprise attack. More advanced automation and autonomous systems could also help to improve the security, efficiency, and resiliency of military communications and command-and-control systems, which are subject to increased disruption in the cyber, electromagnetic, and physical domains. This greater resilience could reduce the vulnerability of nuclear communications and command-and-control systems to disruption.

Conventional Advantages and Nuclear Escalation

Autonomous systems could also influence the potential for nuclear escalation indirectly from the potential for robotic and autonomous systems to give large conventional military advantages to their adopters.

---

[129] Kania, "Battlefield Singularity," 16.
[130] Some types of automation might be beneficial when one party does it, but harmful when deployed symmetrically by both sides, as is the case with many military actions viewed in the context of nuclear stability. One example could be the role of automation in accelerating actions. This could be beneficial in enhancing deterrence and buying additional time for decision-makers. If both sides were to use automation to react at "machine speed," however, the net result could be harmful. Schelling's discussion of the "premium on haste" may also be useful here. Thomas C. Schelling, *Arms and Influence* (New Haven,: Yale University Press, 1966), 227.
[131] Kelsey D. Atherton. "Targeting the future of the DoD's controversial Project Maven initiative," *C4ISRnet*, July 27, 2018, https://www.c4isrnet.com/it-networks/2018/07/27/targeting-the-future-of-the-dods-controversial-project-maven-initiative/

For instance, the Eisenhower doctrine of massive retaliation sought to "knock out [the Soviet] SAC first" once major war appeared imminent, thereby enabling the United States to use nuclear weapons to offset the Warsaw Pact's numerical advantage in conventional forces.[132] Some fear that Russia, now in a position of conventional inferiority relative to the United States, might escalate to nuclear war more quickly during a conflict for similar reasons.[133] Concern about the way that U.S. conventional superiority might encourage nuclear-armed adversaries to escalate in a crisis led the U.S. Department of Defense, in the 2014 Quadrennial Defense Review, to discuss U.S. nuclear deterrence as critical, in part, for "communicating to potential nuclear-armed adversaries that they cannot escalate their way out of failed conventional aggression."[134]

If a nation deploys autonomous systems in a manner that significantly tilts the conventional military relationship in its favor, a nuclear-armed adversary may feel increased incentives to resort to nuclear use, or the threat thereof, to avoid military defeat.[135] For instance, the advent of autonomous swarming forces with greater speed and coordination than human-centered systems could heighten conventional military imbalances in key regions, such as Europe or Asia. This possibility has already attracted the attention of prominent Russian strategists, who write that robotic systems could one day "yield results comparable to the battlefield efficiency of nuclear weapons."[136] If autonomous military systems provide an actor like the United States – or others in the future – a decisive conventional military advantage versus a nuclear-armed adversary, that adversary may be incentivized to threaten or even use nuclear weapons to defeat aggression or coerce an end to the conflict. Russia is not the only country that has indicated a possible willingness to use limited nuclear strikes to end a conventional war that it is losing on favorable terms. Pakistan has indicated likewise.[137] And China may be considering a similar doctrinal innovation.[138]

---

[132] "Memorandum of Discussion at the 227th Meeting of the National Security Council, Washington, August 4, 1955," December 3, 1954, *Foreign Relations of the United States*, 1952-54, Vol. 2, Part 1, https://history.state.gov/historicaldocuments/frus1952-54v02p1/d138; and "Memorandum of Discussion at the 257th Meeting of the National Security Council, Washington, August 4, 1955," August 4, 1955, *Foreign Relations of the United States*, 1955-57, Vol. 19, https://history.state.gov/historicaldocuments/frus1955-57v19/d30. For a broader review of the Eisenhower administration's nuclear doctrine, see Michael Gerson in *Strategic Stability: Contending Interpretations*, 12-13. See also John Lewis Gaddis, *Strategies of Containment: A Critical Appraisal of Postwar American National Security Policy* (New York: Oxford University Press, 1982).
[133] U.S. Department of Defense, "Nuclear Posture Review 2018." 30. For a definitive review of the Russian military-theoretical literature on this subject, see Dave Johnson, "Russia's Conventional Precision Strike Capabilities, Regional Crises, and Nuclear Thresholds," Livermore Papers on Global Security No. 3 (Lawrence Livermore National Laboratory, February 2018).
[134] Department of Defense, "Quadrennial Defense Review," 2014, http://archive.defense.gov/pubs/2014_Quadrennial_Defense_Review.pdf, 13
[135] Elbridge Colby, "Nuclear Weapons in the Third Offset Strategy: Avoiding a Nuclear Blind Spot in the Pentagon's New Initiative," Beyond Offset Series (Center for a New American Security, February 2015), https://s3.amazonaws.com/files.cnas.org/documents/Nuclear-Weapons-in-the-3rd-Offset-Strategy.pdf.
[136] V.N. Gorbunov and S.A. Bogdanov, "Armed Confrontation in the 21st Century," *Military Thought*, 1 (2009).

It is also possible, however, that robotics and autonomous systems narrow the gap between nuclear powers, decreasing reliance on nuclear weapons. Given that the key driver of robotics and AI technology is the commercial sector – and that robotic technologies to-date have rapidly diffused – AI could end up being more of a net leveler among actors from a balance of power perspective. More sophisticated actors would still have access to more capable military systems, but the relatively low barriers to entry for AI and autonomous systems compared to other military-specific technologies such as stealth or fighter jet engines, means that less capable actors would gain in relative power. If applications of AI serve to narrow conventional military gaps, the result could actually decrease the reliance that some nuclear powers place on nuclear weapons, because they would feel more capable of defending themselves conventionally.

Unintentional Escalation

The use of autonomous systems for conventional military operations also could increase the risk of accidental, inadvertent, or catalytic escalation at the conventional level, with implications for nuclear stability. For instance, if nations developed lethal autonomous weapons that could choose their own targets without a human in the loop, they could potentially be at risk for accidents, miscalculation, or unanticipated interactions with the environment or adversary systems. During a crisis, such unauthorized interactions could pose risks to escalation management.[139]

A more direct and worrisome consequence would be accidents that led to unintended attacks against adversary nuclear or dual-use command-and-control systems. For instance, autonomous kinetic-strike systems might inadvertently hit elements of an enemy theater command-and-control system that is linked to its strategic command-and-control system. Or, autonomous cyber capabilities – active defensive or offensive – might "stumble into" sensitive areas in an adversary's strategic military networks during a crisis or conflict, thereby inciting escalation.[140]

---

[137] Feroz Hassan Khan, "Going Tactical: Pakistan's Nuclear Posture and Implications for Stability," Proliferation Papers 53 (Institut français des relations internationals, September 2015), https://www.ifri.org/sites/default/files/atoms/files/pp53khan_0.pdf; and Toby Dalton and George Percovich, "India's Nuclear Options and Escalation Dominance" (Carnegie Endowment for International Peace, May 2016), http://carnegieendowment.org/files/CP_273_India_Nuclear_Final.pdf.
[138] Dennis C. Blair and Caitlin Talmadge, "Would China Go Nuclear?" *Foreign Affairs* (January/February 2019); Caitlin Talmadge, "Beijing's Nuclear Option," *Foreign Affairs* (November/December 2018); and Elbridge Colby, "If You Want Peace, Prepare for Nuclear War" (November/December 2018).
[139] Scharre, "Autonomous Weapons and Operational Risk." Also see Geist and Lohn, "How Might Artificial Intelligence Affect the Risk of Nuclear War."
[140] Technology for Global Stability, "Artificial Intelligence and the Military: Forever Altering Strategic Stability," https://www.tech4gs.org/uploads/1/1/1/5/111521085/ai_and_the_military_forever_altering_strategic_stability__t4gs_research_paper.pdf.. Also see Pavel Sharikov, "Artificial intelligence, cyberattack, and nuclear weapons – A dangerous combination," *Bulletin of the Atomic Scientists*, November 1, 2018, https://thebulletin.org/2018/11/artificial-intelligence-cyberattack-and-nuclear-weapons-a-dangerous-combination/, and Geist and Lohn, "How Might Artificial Intelligence Affect the Risk of Nuclear War."

Catalytic (third party) escalation at the conventional level also poses a risk for nuclear stability. A third party observing rising tensions between two of its own nuclear-armed adversaries may be incentivized to instigate a crisis or conflict, in the expectation that it would benefit by turning its adversaries on one another. Autonomous systems may offer a particularly viable option for these actors. For instance, if the two states are engaged in a limited conventional war, a third party could hack an autonomous system fielded by one of the parties and use it to expand the scope of the conflict, for instance, by targeting previously untargeted theater or strategic assets. The risk of catalytic escalation would be exacerbated by actors' unfamiliarity with autonomous systems, the increased speed of military activity in an era of AI, and more traditional drivers of crisis escalation, to include imperfect information about an adversary's intentions and capabilities. Third party actors may be able to exploit these crisis dynamics to preserve their anonymity. For instance, they could redirect targeting by an autonomous system during a period of intensified fighting. This could leave the targeted actor with very little time to evaluate the likely source or intent of escalated attacks on its theater or strategic assets, thereby increasing the likelihood of counter-escalation.

Escalation in each of the scenarios described above would be driven, at least at first, by conventional systems. Unwanted escalation below the nuclear threshold, however, could still increase the risk of crises or conflicts getting out of hand by dragging nations further down the slippery slope from crisis to conflict or conventional conflict to nuclear war.

## Conclusion

Automation is in many ways a double-edged sword. When used for predictable tasks or to respond to known events, automation can be a boon, increasing safety and reliability. Automated systems are often "brittle," however. They can perform exceptionally well under known conditions, but when pushed outside the bounds of their operating environment, they can often fail badly. Moreover, as automated systems become more complex, it can be increasingly difficult for users and even designers to reliably predict when systems may fail.

In spite of these risks, automation overall has tremendous potential advantages for improving safety and reliability in a variety of applications, including nuclear operations. Automated systems can execute routine tasks more precisely and reliably than humans and are not subject to fatigue or distraction. This is why automation is undoubtedly a major factor in improvements in airline safety.[141] Self-driving cars likewise have tremendous potential to reduce automobile accidents, which kill over 30,000 people annually in the United States alone.[142]

---

[141] For instance, the introductions of "glass cockpits," flight management systems, and GPS has led to significant decreases in Controlled Flight Into Terrain Accident Rates. Flight envelope protection has also led to a substantial decrease in Loss of Control In-Flight accident rates. See Airbus, "Commercial Aviation Accidents: 1958-2014 – A Statistical Analysis," http://asndata.aviation-safety.net/industry-reports/Airbus-Commercial-Aviation-Accidents-1958-2014.pdf. At the same time, however, automation does have its limits. Maria Konnikova, "The Hazards of Going on Autopilot," *The New Yorker*, September 4 2014, http://www.newyorker.com/science/maria-konnikova/hazards-automation.

[142] "Fatality Facts," Insurance Institute for Highway Safety, http://www.iihs.org/iihs/topics/t/general-statistics/fatalityfacts/state-by-state-overview.

Automation could thus complicate human decision-making in crises in surprising and perhaps counterintuitive ways. As Michael Carl Haas of ETH Zurich has argued, "[T]he artificial intelligence (AI) that governs the behavior of autonomous systems during their operational employment would become an additional actor participating in the crisis, though one who is tightly constrained by a set of algorithms and mission objectives."

A key factor mediating the interaction between autonomous systems and nuclear security will involve psychology – potentially the way automation bias and other cognitive biases could shape decision making.[143] The same contradictions that pose challenges for nuclear stability overall remain present in decisions about whether or not to automate. A major factor in crises is not only how automation influences the user's decision-making, but also how it affects the adversary's calculus. In theory, automation could be used to enhance deterrence by effectively tying one's hands, for instance, by communicating that any attack on a nation's homeland defense systems would trigger nuclear escalation. Making such a threat credible will be difficult, however.

The analysis in this paper demonstrates that it remains to be seen whether automation has a net positive or negative effect on nuclear stability, but the potential risks are large. Whether it does could depend on the specific application – both where it falls in the nuclear enterprise but also how it is implemented in terms of design, human-machine interfaces, training, and operator culture. Like many things relating to nuclear stability, the value or harm in automation might also appear differently when considered from the net perspective of both sides. How automation is perceived also matters. If policymakers view automation as unpredictable and unreliable and therefore its use induces caution, then countries might be less willing to engage in brinkmanship. On the other hand, if nations viewed automation as more safe and reliable than it actually was, then it could lead policymakers to underestimate the chances of accidents or miscalculation and take risks they do not understand.

As in other military and civilian applications, greater automation is likely to creep its way into nuclear operations over time, especially as nations modernize their nuclear forces. Additionally, the increasing use of automation and autonomous systems in other aspects of military operations could affect nuclear stability. Many of these applications of automation could potentially enhance nuclear stability, but others could undermine it. In the worst cases, states may not fully understand the risks posed by automation leading them to misjudge adversaries' responses or potentially even how their own forces might behave in a crisis. Concepts such as retaining a human "in the loop," while well-intentioned, are not a panacea against accidents. Automation bias can pose insidious risks that do not manifest until an accident occurs. Future research should think through, in greater details, the nuclear escalation risks posed by AI and autonomous systems in order to anticipate these challenges and, when possible, take the necessary steps to avert them.

Technology is not destiny. The rapid progress of AI and automation in the commercial sector opens up opportunities for militaries, but militaries have a choice about how they incorporate automation into their forces. Some forms of automation could increase reliability and surety in nuclear operations, strengthening stability, while other forms could increase accident risk or create perverse incentives, undermining stability. As in other aspects of nuclear stability, second- and third-order consequences must be understood. Actions that appear beneficial can sometimes have counterintuitive consequences, especially when accounting for an adversary's decision calculus. When modernizing nuclear arsenals, policymakers should aim to use automation to decrease the risk of accidents and false alarms and increase human control over nuclear operations.

---

[143] Kenneth Payne. *Strategy, Evolution, and War: From Apes to Artificial Intelligence* (Washington, DC: Georgetown University Press, 2018).